\documentclass[letterpaper,english,reprint,aps,prb,showkeys]{revtex4-1}
\usepackage[T1]{fontenc}
\usepackage[latin9]{inputenc}
\usepackage{array}
\usepackage{booktabs}
\usepackage{amsmath}
\usepackage{amssymb}
\usepackage{graphicx}
\usepackage{esint}

\makeatletter

\providecommand{\tabularnewline}{\\}

\RequirePackage{fix-cm}

\usepackage{physics}

\def\ignorecitefornumbering#1{%
     \begingroup
         \@fileswfalse
         #1
    \endgroup
}

\usepackage{babel}

\usepackage[labelsep=period]{caption}
\renewcommand\toprule{\hline\hline}
\renewcommand\bottomrule{\hline\hline}

\makeatother

\usepackage{babel}
\begin{document}

\preprint{APS/123-QED}

\title{Thermodynamics of adiabatic melting of solid \textsuperscript{4}He
in liquid \textsuperscript{3}He}

\author{T. S. Riekki}
\email{tapio.riekki@aalto.fi}

\selectlanguage{english}%

\affiliation{Aalto University - School of Science, Low temperature laboratory,
P.O. BOX 15100 FI-00076 AALTO, Finland }

\author{A. P. Sebedash}

\affiliation{P. L. Kapitza Institute for Physical Problems RAS, Kosygina 2, 119334
Moscow, Russia }

\author{J. T. Tuoriniemi}

\affiliation{Aalto University - School of Science, Low temperature laboratory
P.O. BOX 15100 FI-00076 AALTO, Finland }

\date{\today}
\begin{abstract}
In the cooling concept by adiabatic melting, solid \textsuperscript{4}He
is converted to liquid and mixed with \textsuperscript{3}He to produce
cooling power directly in the liquid phase. This method overcomes
the thermal boundary resistance that conventionally limits the lowest
available temperatures in the helium fluids, and hence makes it possible
to reach for the temperatures significantly below $100\,\mathrm{\mu K}$.
In this paper we focus on the thermodynamics of the melting process,
and examine the factors affecting the lowest temperatures achievable.
We show that the amount of \textsuperscript{3}He\textendash \textsuperscript{4}He
mixture in the initial state, before the melting, can substantially
lift the final temperature, as its normal Fermi fluid entropy will
remain relatively large compared to the entropy of superfluid \textsuperscript{3}He.
We present the collection of formulas and parameters in order to work
out the thermodynamics of the process at very low temperatures, study
the heat capacity and entropy of the system with different liquid
\textsuperscript{3}He, mixture, and solid \textsuperscript{4}He
contents, and use them to estimate the lowest temperatures achievable
by the melting process, as well as compare our calculations to the
experimental saturated \textsuperscript{3}He\textendash \textsuperscript{4}He
mixture crystallization pressure data. Realistic expectations in the
execution of the actual experiment are considered. Further, we study
the cooling power of the process, and find the coefficient connecting
the melting rate of solid \textsuperscript{4}He to the dilution rate
of \textsuperscript{3}He.

\begin{description}
\item [{PACS~numbers}] 67.60.G-, 67.30.ef, 67.80.B- {\small \par}
\end{description}
\end{abstract}



\maketitle

\section{Introduction\label{sec:Introduction}}

One of the persistent great problems in the field of low-temperature
physics is the search for superfluidity of \textsuperscript{3}He
diluted by \textsuperscript{4}He. Superfluidity of pure \textsuperscript{4}He
was discovered in the late 1930s\citep{KAPITZA1938,ALLEN1938} at
around $2\,\mathrm{K}$, but the superfluidity of pure \textsuperscript{3}He
was observed not until three decades later, at three orders of magnitude
lower temperature\citep{Osheroff1972,Osheroff1972a}. To achieve that,
completely new cooling methods had to be developed. In the quest for
superfluidity of \textsuperscript{3}He in \textsuperscript{3}He\textendash \textsuperscript{4}He
mixture, we are in a similar situation: we have exhausted the search
space available using the present cooling techniques, and hence a
new approach is needed. The target temperatures are below $100\,\mathrm{\mu K}$,
where a BCS-type superfluid transition is expected to occur between
weakly interacting \textsuperscript{3}He atoms in the isotope mixture\citep{Bardeen1967,Effective_3He_interactions,AL-SUGHEIR2006,Sandouqa2011}.
Such a system would be a unique dense double-superfluid ensemble consisting
of fermionic \textsuperscript{3}He and bosonic \textsuperscript{4}He
superfluids. In sparse ultracold atomic gases, this kind of mixture
superfluid phase has already been observed\citep{Bose_Fermi_superfluids,Roy2017}.

Adiabatic melting of solid \textsuperscript{4}He followed by its
mixing with \textsuperscript{3}He is one novel cooling technique
proposed to achieve this temperature range. The major advantage of
the method is that it bypasses the rapidly increasing thermal boundary
resistance that limits the lowest temperatures available with external
cooling methods, such as adiabatic nuclear demagnetization. Even as
the walls of a helium container can be cooled to tens of microkelvins
range, the liquid inside will remain at an elevated temperature due
to the poor thermal coupling across the thermal boundary resistance
bottleneck. No matter how small the heat load to the liquid is, the
cooling power across the bottleneck will struggle to overcome it when
temperature is low enough. The lowest directly measured temperature
in \textsuperscript{3}He\textendash \textsuperscript{4}He mixture
was $97\,\mathrm{\mu K}$ reported by Oh \emph{et al.} \citep{Oh1994},
achieved in an experimental volume with about $4000\,\mathrm{m^{2}}$
surface area cooled by a two-stage nuclear demagnetization cryostat.

Since the cooling by adiabatic melting takes place directly in liquid
helium, the thermal boundary resistance is no longer the main factor
limiting the final temperature. In this technique, first, a system
of solid \textsuperscript{4}He and liquid \textsuperscript{3}He
is precooled to as low temperature as possible with an external cooling
method. Then, as the solid is melted, it releases liquid \textsuperscript{4}He
which will be mixed with \textsuperscript{3}He producing cooling
due to the latent heat of mixing. The principle of operation is somewhat
similar to a conventional dilution refrigerator, the difference being
that the adiabatic melting method is not continuous, and takes place
at an elevated pressure. \citep{Adiabatic_Melting,Sebedash_QFS,Sebedash1997}

The realization of the adiabatic melting experiment is quite a technical
challenge \citep{Sebedash_QFS}. In this paper, however, we focus
on the thermodynamic aspects: the success of the melting process depends
on the initial conditions, and the proper execution. The final temperature
ultimately achievable by this method is determined by the initial
contents of the experimental cell, which should have as little entropy
as possible to begin with. An ideal initial state would contain only
solid pure \textsuperscript{4}He which has negligibly small entropy,
and pure superfluid \textsuperscript{3}He. Below the superfluid transition
temperature, its entropy decreases exponentially with temperature,
so that even a small reduction in the starting temperature can significantly
diminish the entropy content of the total system. In actual cases,
however, there is always some \textsuperscript{3}He\textendash \textsuperscript{4}He
mixture present, and already in quite small quantities its contribution
easily dominates the total initial entropy.

We begin this paper by presenting the formulas and parameters to calculate
the heat capacity and entropy of the pure \textsuperscript{3}He\textendash mixture\textendash solid
\textsuperscript{4}He-system at \textsuperscript{4}He crystallization
pressure, and then we examine the final temperatures that can be reached
with different starting conditions. A figure of merit is the cooling
factor, \emph{i.e.}, the ratio between the initial and the final temperature.
We will also discuss the cooling power due to the melting/mixing process,
which is proportional to the phase transfer rate of \textsuperscript{3}He,
which, in turn, is related to the rate at which the \textsuperscript{4}He
crystal is melted. Finally, we will compare the temperature dependence
of the crystallization pressure deduced from our calculations to the
experimentally obtained values. 

\section{Heat capacity and entropy\label{sec:Entropy-and-heat-capacity}}

We concentrate on the low temperature properties (mostly below  $10\,\mathrm{mK}$)
of phase-separated \textsuperscript{3}He\textendash \textsuperscript{4}He
mixture at its crystallization pressure $P_{\mathrm{C}}=2.564\,\mathrm{MPa}$
\citep{Pentti_Thermometry}. Our system thus consists of liquid rich
and dilute \textsuperscript{3}He phases, as well as solid \textsuperscript{4}He
phase. Under these conditions the \textsuperscript{3}He rich phase
is pure \textsuperscript{3}He, while the \textsuperscript{3}He dilute
phase contains a certain amount of \textsuperscript{3}He down to
the zero-temperature limit. This finite solubility is the basis of
not only the conventional dilution refrigerator, but also the adiabatic
melting method. Superfluid \textsuperscript{4}He of the mixture phase
is basically in its quantum mechanical ground state and it acts as
an inert background for the \textsuperscript{3}He quasiparticles
affecting on their effective mass \citep{Dobbs2001}. Meanwhile, the
solid phase can be assumed to be pure \textsuperscript{4}He, provided
that the crystal was grown at sufficiently low temperature ($\ll50\,\mathrm{mK}$)
\citep{Pantalei2010,Balibar2000,Balibar2002}.

The only free thermodynamic parameter of the system is temperature
$T$, as the solid \textsuperscript{4}He phase fixes the pressure
to the crystallization pressure, and the presence of the rich \textsuperscript{3}He
phase ensures that the dilute \textsuperscript{3}He\textendash \textsuperscript{4}He
mixture remains at its saturation concentration $x=8.12\%$ \citep{Pentti_etal_solubility}.
The system is thus a univariant three-phase system.

The Fermi systems in question, \textsuperscript{3}He in the rich
or dilute phase, are deep in the degenerate state so that the normal
fluid heat capacity is directly proportional to the temperature $C\propto T$.
At the superfluid transition temperature of the pure \textsuperscript{3}He
($T_{\mathrm{c}}$), its heat capacity suddenly increases and then
drops exponentially towards lower temperatures. The isotope mixture,
on the other hand, maintains the linear temperature dependence down
to much lower temperatures, so that even a very small amount of \textsuperscript{3}He\textendash \textsuperscript{4}He
mixture will eventually dominate the heat capacity of the entire system.
Compared to that, we can ignore the phonon contributions to the heat
capacity. This applies to all phases present, and in particular the
heat capacity of the solid \textsuperscript{4}He can thus be approximated
as zero. Further, we assume that the molar volumes of all phases remain
constant. 

For the $T_{\mathrm{c}}$ of pure \textsuperscript{3}He at the \textsuperscript{3}He\textendash \textsuperscript{4}He
mixture crystallization pressure, we use the value $2.6\,\mathrm{mK}$
given by Pentti \emph{et al.} \citep{Pentti_Thermometry}. This is
about 10\% higher than the transition temperature suggested by the
provisional PLTS-2000 \citep{Rusby2002} temperature scale, but it
is consistent with other characteristic \textsuperscript{3}He temperatures,
such as A-B and Néel transition temperatures, carefully determined
at our cryostat during other experiments \citep{Manninen2016}. The
precise value of the $T_{\mathrm{c}}$ is not critical to the most
of the analysis presented in this paper however, as the heat capacity
and entropy can be given with respect to their value at the $T_{\mathrm{c}}$.

The values for the parameters used in the following calculations are
listed in Table~\ref{tab:Parameters}.

The heat capacity for $n$ moles of degenerate Fermi fluid is given
by\citep{Lifshitz1980} 
\begin{equation}
\frac{C}{nR}=\frac{\pi^{2}}{2}\frac{T}{T_{\mathrm{F}}},\label{eq:C1}
\end{equation}
where $R$ is the molar gas constant, and $T_{\mathrm{F}}$ the Fermi
temperature. The heat capacity of normal fluid pure \textsuperscript{3}He
is usually expressed with the Sommerfeld constant $\gamma=\pi^{2}\left(2T_{\mathrm{F}}\right)^{-1}$
as
\begin{equation}
\frac{C_{3}\left(T>T_{\mathrm{c}}\right)}{n_{3}R}=\gamma T,\label{eq:C3n}
\end{equation}
where $n_{3}$ is the amount of \textsuperscript{3}He in the pure
phase. By interpolating the data given by Greywall \citep{Greywall1986},
we can find $4.14\,\mathrm{K^{-1}}$ for the $\gamma$-parameter at
the $P_{\mathrm{C}}$. But the temperature scales used by Greywall
and us differ by about 10\%, as manifested by the different superfluid
transition temperature values used by us ($T_{\mathrm{c}}=2.6\,\mathrm{mK}$)
and Greywall ($T_{\mathrm{c,Gw}}=2.37\,\mathrm{mK}$). Now, since
$\gamma=C/T=\mathrm{dQ}/\left(T\mathrm{d}T\right)$, we need to scale
the above value by $\left(T_{\mathrm{c},\mathrm{Gw}}/T_{\mathrm{c}}\right)^{2}$
to maintain consistency, giving us $\gamma=3.44\,\mathrm{K}^{-1}$.
The value obtained this way is close to the coefficient interpolated
from the data by Alvesalo \emph{et al.} \citep{Alvesalo1981}, approximately
$3.2\,\mathrm{K^{-1}}$ at the $P_{\mathrm{C}}$. Although the about
20\% margin in the $\gamma$-parameter is rather inconvenient, it
is not unheard of. As already pointed out by Greywall \citep{Greywall1986},
this magnitude of discrepancy in the $\gamma$-values obtained by
various experimental groups can be attributed to the difference in
their temperature scales.\setlength{\extrarowheight}{0.2cm} 
\begin{table*}[t]
\center%
\begin{tabular}{l>{\raggedright}p{0.4\textwidth}>{\centering}p{0.1\textwidth}>{\centering}p{0.2\textwidth}>{\centering}p{0.2\textwidth}c}
\toprule 
 & parameter & symbol & value & Ref. & \tabularnewline
\midrule
 & \textsuperscript{3}He\textendash \textsuperscript{4}He mixture crystallization
pressure & $P_{\mathrm{C}}$ & 2.564 MPa & [\ignorecitefornumbering{\onlinecite{Pentti_Thermometry}}] & \tabularnewline
 & \textsuperscript{3}He Sommerfeld constant$^{\ddagger}$ & $\gamma$ & 3.44 K\textsuperscript{-1} & [\ignorecitefornumbering{\onlinecite{Greywall1986}}] & \tabularnewline
 & \textsuperscript{3}He superfluid transition temperature & $T_{\mathrm{c}}$ & 2.6 mK & [\ignorecitefornumbering{\onlinecite{Pentti_Thermometry}}] & \tabularnewline
 & \textsuperscript{3}He effective mass in \textsuperscript{3}He\textendash \textsuperscript{4}He
mixture & $m^{*}/m_{3}$ & 3.32 & [\ignorecitefornumbering{\onlinecite{Effective_3He_interactions}}] & \tabularnewline
 & saturation concentration & $x$ & 8.12\% & [\ignorecitefornumbering{\onlinecite{Pentti_etal_solubility}}] & \tabularnewline
 & BBP-parameter & $\alpha$ & 0.164 & [\ignorecitefornumbering{\onlinecite{Watson_Reppy_Richardson}}] & \tabularnewline
 & liquid \textsuperscript{4}He molar volume & $V_{4,\mathrm{l}}$ & 23.16 cm\textsuperscript{3}/mol & [\ignorecitefornumbering{\onlinecite{Tanaka2000}}] & \tabularnewline
 & solid \textsuperscript{4}He molar volume & $V_{4,\mathrm{s}}$ & 20.97 cm\textsuperscript{3}/mol & [\ignorecitefornumbering{\onlinecite{Driessen1986}}] & \tabularnewline
 & liquid \textsuperscript{3}He molar volume & $V_{3}$ & 26.76 cm\textsuperscript{3}/mol & [\ignorecitefornumbering{\onlinecite{Kollar2000a}}] & \tabularnewline
 & liquid \textsuperscript{3}He\textendash \textsuperscript{4}He mixture
molar volume & $V_{\mathrm{m}}$ & 23.47 cm\textsuperscript{3}/mol & Eq.~\eqref{eq:mix-molar-vol} & \tabularnewline
 & \textsuperscript{3}He\textendash \textsuperscript{4}He mixture Fermi
temperature & $T_{\mathrm{F,m}}$ & 0.378 K & Eq.~\eqref{eq:mix-Fermi} & \tabularnewline
 & \textsuperscript{3}He Fermi temperature & $T_{\mathrm{F,3}}$ & 1.43 K & $\overset{}{=\pi^{2}(2\gamma)^{-1}}$ & \tabularnewline
 & superfluid \textsuperscript{3}He energy gap & $\Delta_{0}$ & $1.91T_{\mathrm{c}}$ & [\ignorecitefornumbering{\onlinecite{Todoschenko2002}},\ignorecitefornumbering{\onlinecite{Serene1983}}] & \tabularnewline
 & 1st fitting parameter & $A$ & 8.242 & Eq.~\eqref{eq:C3s} & \tabularnewline
 & 2nd fitting parameter & $B$ & 11.22 & Eq.~\eqref{eq:C3s} & \tabularnewline
\bottomrule
\end{tabular}

\caption{Values of the parameters used in our calculations at saturated \protect\textsuperscript{3}He\textendash \protect\textsuperscript{4}He
mixture crystallization pressure. $V_{\mathrm{m}}$, $T_{\mathrm{F,m}}$,
and $T_{\mathrm{F,3}}$ were calculated using the other listed parameters,
while $A$ and $B$ are fitting parameters for the pure \protect\textsuperscript{3}He
heat capacity below the $T_{\mathrm{c}}$. $^{\ddagger}$This value
was scaled due to the difference in the temperature scales between
Ref.~[\onlinecite{Greywall1986}] and us (see text). \label{tab:Parameters}}
\end{table*}

In order to describe the behavior of pure \textsuperscript{3}He below
the superfluid transition temperature $T_{\mathrm{c}}$ with a single
smooth function over the entire temperature range, we use the expression
\begin{equation}
C_{3}\left(T\leq T_{\mathrm{c}}\right)=\gamma T_{\mathrm{c}}\left[\left(\frac{A}{\widetilde{T}}+B\widetilde{T}^{2}\right)\exp\left(-\frac{\Delta_{0}}{T}\right)\right],\label{eq:C3s}
\end{equation}
where $\widetilde{T}=T/T_{\mathrm{c}}$ is the reduced temperature,
$A$ and $B$ are fitting parameters, and $\Delta_{0}=1.91T_{\mathrm{c}}$
is the superfluid \textsuperscript{3}He energy gap at the zero-temperature
limit, taken as average of the values given by Refs.~[\onlinecite{Todoschenko2002}]
and [\onlinecite{Serene1983}]. The fit was made against the normalized
heat capacity data by Greywall \citep{Greywall1986}. 

The heat capacity of \textsuperscript{3}He\textendash \textsuperscript{4}He
mixture is given by Eq.~\eqref{eq:C1}, using the Fermi temperature
of the mixture as \citep{Lifshitz1980,Dobbs2001}
\begin{equation}
T_{\mathrm{F,m}}=\frac{\hbar^{2}}{2m^{*}k_{\mathrm{B}}}\left(\frac{3\pi^{2}N_{\mathrm{A}}x}{V_{\mathrm{m}}}\right)^{2/3},\label{eq:mix-Fermi}
\end{equation}
where $x=8.12\%$ \citep{Pentti_etal_solubility} is the saturation
concentration of the mixture, and $\hbar$, $k_{\mathrm{B}}$, and
$N_{\mathrm{A}}$ are the reduced Planck constant, Boltzmann constant,
and Avogadro constant, respectively. The effective mass of \textsuperscript{3}He
atom $m^{*}=3.32m_{3}$ ($m_{3}=3.0160293\,\mathrm{u}$ is the bare
\textsuperscript{3}He mass) was calculated using the quasiparticle
interaction potential from Ref.~[\onlinecite{Effective_3He_interactions}]
at the saturation concentration. Next, the molar volume of the mixture
$V_{\mathrm{m}}$ is evaluated from \citep{Bardeen1967}
\begin{equation}
V_{\mathbf{\mathrm{m}}}=V_{\mathrm{4,l}}\left(1+\alpha x\right),\label{eq:mix-molar-vol}
\end{equation}
where $V_{\mathrm{4,l}}=23.16\,\mathrm{cm^{3}/mol}$ \citep{Tanaka2000}
is the molar volume of liquid pure \textsuperscript{4}He, and $\alpha$
is the BBP-parameter that describes the extra volume occupied by the
lighter \textsuperscript{3}He atoms with their larger zero-point
motion. We use the value $\alpha=0.164$ extrapolated from Ref.~[\onlinecite{Watson_Reppy_Richardson}].
With these the mixture molar volume is $V_{\mathrm{m}}=23.47\,\mathrm{cm^{3}/mol}$.
The Fermi temperature of the mixture is thus $T_{\mathrm{F,m}}=0.378\,\mathrm{K}$,
and the heat capacity per mole of \textsuperscript{3}He in the mixture
$n_{\mathrm{m},3}$ becomes 
\begin{equation}
\frac{C_{\mathrm{m,3}}}{n_{\mathrm{m},3}R}=13.05\,\frac{T}{\mathrm{K}}.\label{eq:Cm}
\end{equation}

\begin{figure*}[t]
\center\includegraphics[width=17.8cm]{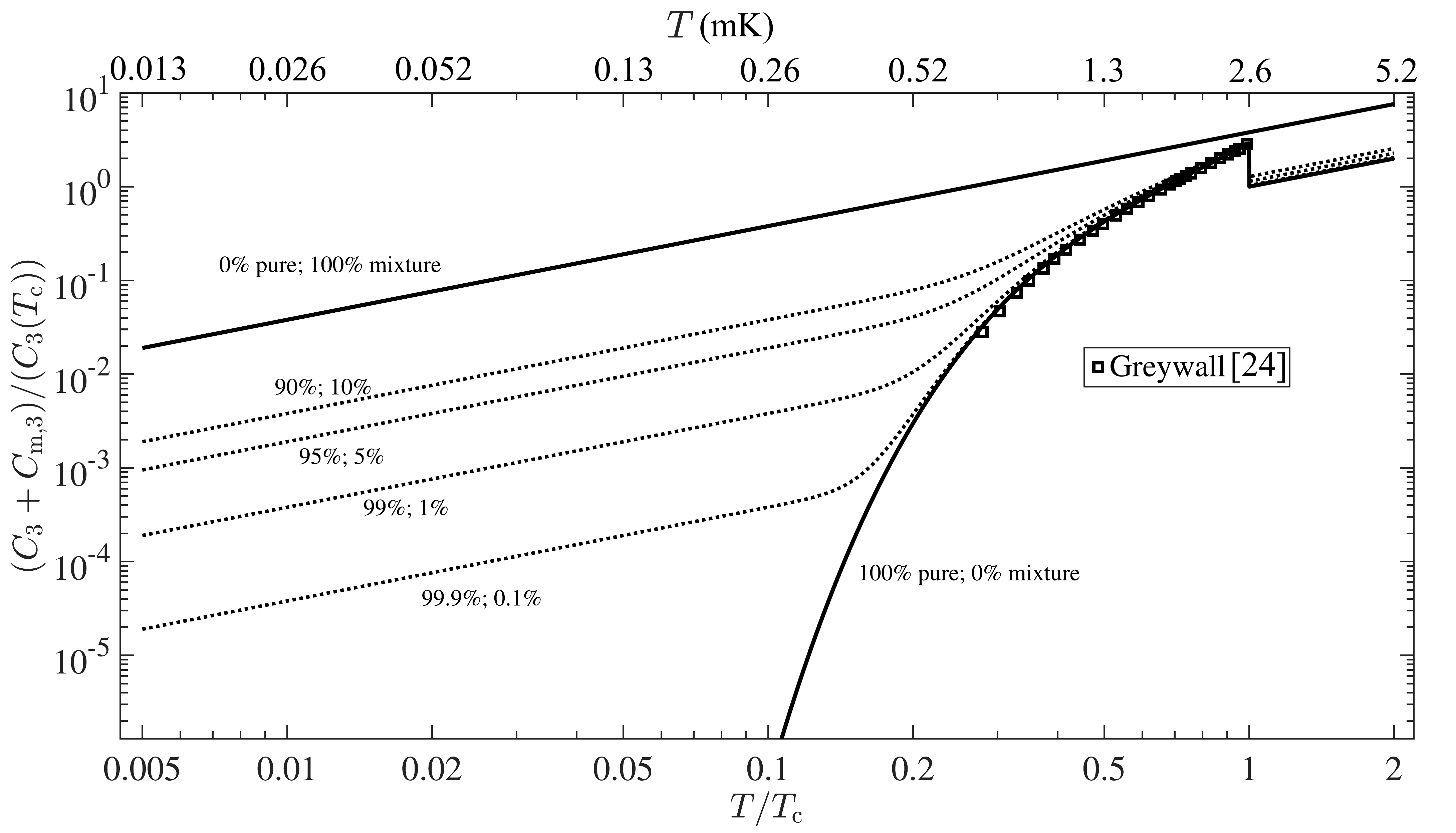}

\caption{Total heat capacity of pure \protect\textsuperscript{3}He ($C_{3}$),
and \protect\textsuperscript{3}He\textendash \protect\textsuperscript{4}He
mixture ($C_{\mathrm{m,3}}$) system per mole of \protect\textsuperscript{3}He
as a function of the temperature below $2T_{\mathrm{c}}$ calculated
from Eqs.~\eqref{eq:C3n},\eqref{eq:C3s}, and \eqref{eq:Cm}. The
heat capacity values are scaled by its value at the pure \protect\textsuperscript{3}He
$T_{\mathrm{\mathrm{c}}}$. Experimental data for pure \protect\textsuperscript{3}He
by Greywall \citep{Greywall1986} are shown for comparison. The percentages
tell how the total amount of \protect\textsuperscript{3}He in the
system is split between pure \protect\textsuperscript{3}He phase
and \protect\textsuperscript{3}He\textendash \protect\textsuperscript{4}He
mixture phase.\label{fig:heat-capacity}}
\end{figure*}
\begin{figure*}[t]
\center\includegraphics[width=17.8cm]{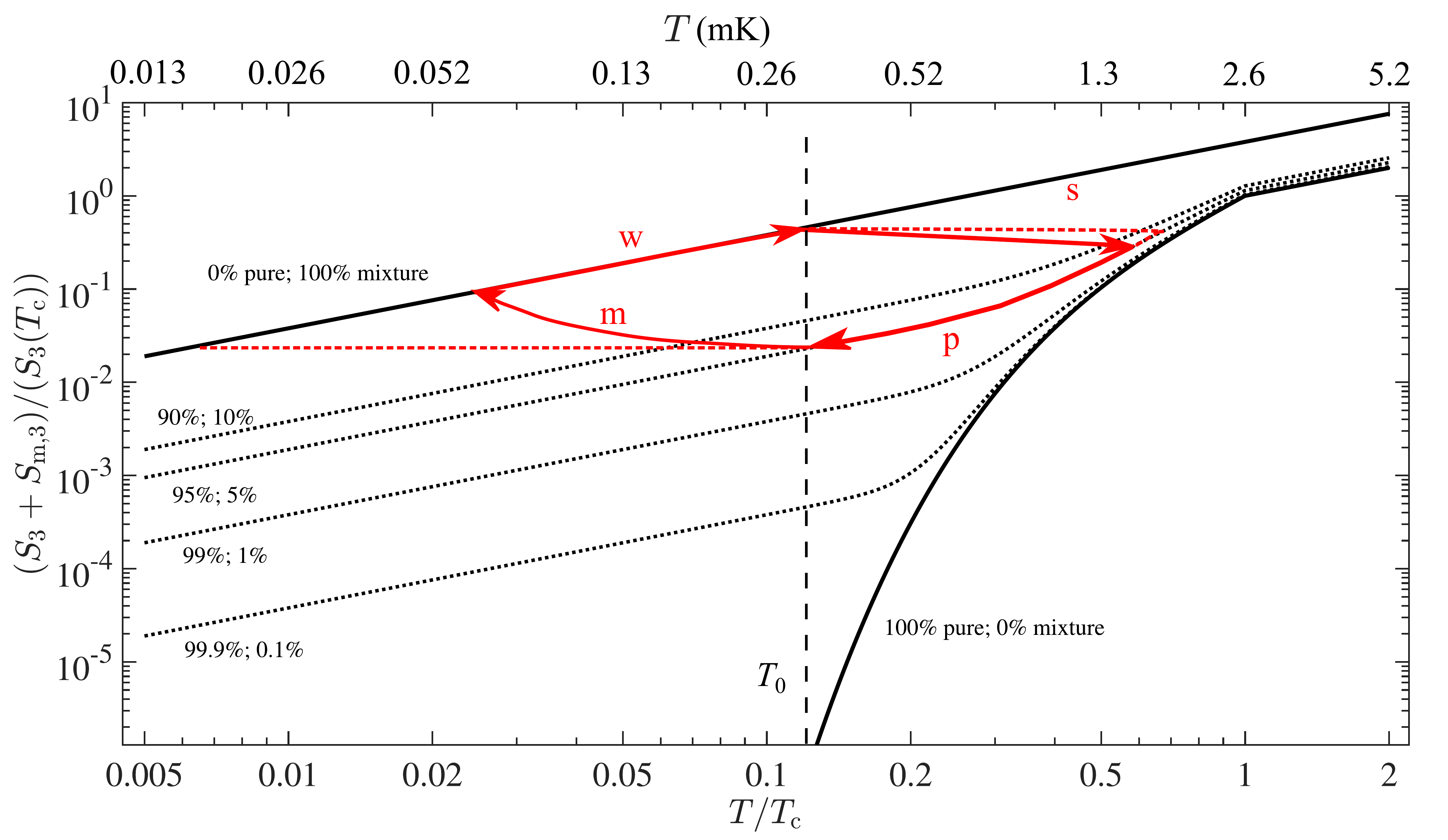}\caption{(color online) Total entropy of pure \protect\textsuperscript{3}He
($S_{3}$), and \protect\textsuperscript{3}He\textendash \protect\textsuperscript{4}He
mixture ($S_{\mathrm{m,3}}$) system per mole of \protect\textsuperscript{3}He
as a function of the temperature below $2T_{\mathrm{c}}$. The entropy
values are scaled by its value at the pure \protect\textsuperscript{3}He
$T_{\mathrm{\mathrm{c}}}$. The red arrows indicate a solidification
(s)\textendash precooling (p)\textendash melting (m)\textendash warm-up
(w) cycle with 5\% of the total \protect\textsuperscript{3}He remaining
in mixture after the solid growth, considered as quite a conservative
value. The precooling temperature is indicated as $T_{0}$. The cycle
was drawn assuming losses during the melting process (m) due to the
heat leak to the experimental cell, while during the solidification
(s) the precooling starts already as the solid is growing. The red
horizontal dashed lines indicate perfectly adiabatic melting and solidification
paths for comparison. The percentages tell how the total amount of
\protect\textsuperscript{3}He in the system is split between pure
\protect\textsuperscript{3}He phase and \protect\textsuperscript{3}He\textendash \protect\textsuperscript{4}He
mixture phase.\label{fig:entropy}}
\end{figure*}

In Fig.~\ref{fig:heat-capacity} we show the heat capacities for
several \textsuperscript{3}He/\textsuperscript{4}He partitions,
starting from a system consisting of pure \textsuperscript{3}He together
with solid \textsuperscript{4}He, and then letting a portion of the
total \textsuperscript{3}He amount to be in the mixture phase so
that the total \textsuperscript{3}He amount of the system remains
constant. Entropies of \textsuperscript{3}He and mixture can then
be calculated from the heat capacity as the integral $S=\intop_{0}^{T}\frac{C}{T'}\mathrm{d}T'$;
in the case of Eq.~\eqref{eq:C3s} numerical integration is required.
They are shown in Fig.~\ref{fig:entropy}.

In an ideal, perfectly adiabatic melting process, where the pure \textsuperscript{3}He
phase shrinks and the mixture phase grows, one moves horizontally
from right to left in diagrams like those of Figs.~\ref{fig:heat-capacity}
and \ref{fig:entropy}, and the final temperature will be determined
by the initial conditions alone. If the initial mixture amount is
vanishingly small, even minor improvements to the precooling conditions
lead to a huge gain in the cooling factor, since the entropy of the
pure superfluid \textsuperscript{3}He phase decreases exponentially.
However, in realistic cases, there is always some small amount of
mixture left. Then, from Fig.~\ref{fig:entropy} we see that precooling
the system to below $0.15T_{\mathrm{c}}$ no longer decreases the
total entropy as rapidly, since even a minuscule mixture amount is
enough to become the main contributor to the total entropy at those
temperatures. $0.15T_{\mathrm{c}}$ is quite a reasonable value for
a decent precooling temperature, as it can be reached using an adiabatic
nuclear demagnetization refrigerator. Figure~\ref{fig:entropy} also
shows an example of an operational cycle of the cooling process, which
begins with the solidification of the \textsuperscript{4}He crystal
(s), followed by the precool along the constant pure \textsuperscript{3}He\textendash mixture
content curve (p). When the set precooling temperature $T_{0}$ is
reached, the melting process is initiated by removing \textsuperscript{4}He
from the cell (m), and at the end one reaches the mixture curve. In
practice, the crystal may not always be completely melted, but as
the remaining undissolved \textsuperscript{3}He has only very small
entropy, this will not greatly affect the final temperature. When
the melting is done, the system may be allowed to warm-up back to
the precooling temperature (w), after which the crystal is regrown.
The cycle in Fig.~\ref{fig:entropy} was drawn by assuming some losses
in the melting process due to the external heat leak to the experimental
cell, which we have taken to be of order $200\,\mathrm{pW}/\mathrm{mole\:^{3}He}$.

\section{Cooling factor\label{sec:Cooling-factor}}

\begin{figure*}[t]
\center\includegraphics[width=17.8cm]{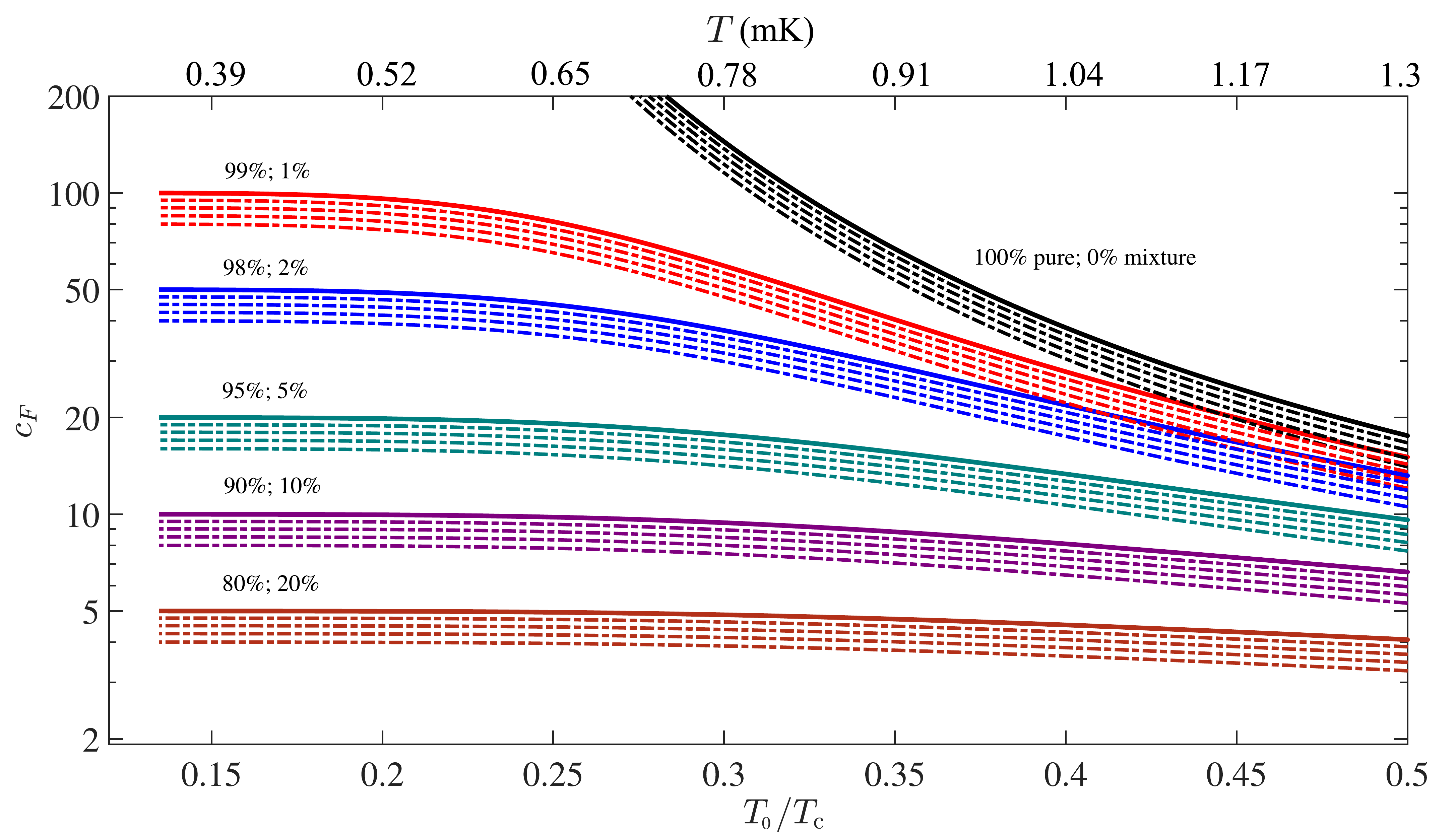}

\caption{(color online) Cooling factors in perfectly adiabatic melting processes,
below $0.5T_{\mathrm{c}}$, as a function of the precooling temperature
$T_{0}$, with different starting conditions from an ideal case with
no mixture phase to one where 20\% of the total \protect\textsuperscript{3}He
is in the mixture phase at the beginning of the melting process. The
solid lines indicate the melting processes that have no pure \protect\textsuperscript{3}He
left in the end, while the dash-dotted lines correspond to incomplete
melting processes ending with 5\%, 10\%, 15\%, and 20\% of the total
\protect\textsuperscript{3}He still in the pure \protect\textsuperscript{3}He
phase. \label{fig:Cooling-factors}}
\end{figure*}

Cooling factor of the adiabatic melting process $c_{\mathrm{F}}$
is defined as the ratio between the temperatures before and after
the melting, $c_{\mathrm{F}}=T_{\mathrm{0}}/T_{\mathrm{final}}$.
Above the pure \textsuperscript{3}He $T_{\mathrm{c}}$, both the
pure and the dilute phase entropy follow the linear temperature dependence,
and hence the cooling factor remains constant. The optimal value for
the cooling factor above the $T_{\mathrm{c}}$ can be evaluated from
Eqs.~\eqref{eq:C3n} and \eqref{eq:Cm} by assuming that the initial
state contains only pure \textsuperscript{3}He and solid \textsuperscript{4}He,
while the final state is exclusively \textsuperscript{3}He\textendash \textsuperscript{4}He
mixture. We get $c_{\mathrm{F}}\left(T>T_{\mathrm{c}}\right)=\pi^{2}/\left(2\gamma T_{\mathrm{F,m}}\right)\approx3.8$.
The conventional dilution refrigerators operate at low pressure with
a cooling factor comparable to this. 

Below the $T_{\mathrm{c}}$, however, the potential cooling factor
increases rapidly as soon as the entropy of pure \textsuperscript{3}He
starts to decrease exponentially. With ideal precooling conditions,
with the system at that stage consisting only of solid \textsuperscript{4}He
and superfluid pure \textsuperscript{3}He, with no mixture phase,
the cooling factor can reach values up to several hundreds as the
precooling temperature approaches $0.15T_{\mathrm{c}}$. The presence
of the mixture phase in the initial state severely limits the possible
cooling factors, as shown in Fig.~\ref{fig:Cooling-factors}. If
even 1\% of the total amount of pure \textsuperscript{3}He remains
in the mixture at the beginning, the cooling factor levels out at
around 100. Further precooling will not help increase it, as the entropy
of the entire system is now dominated by the entropy of the mixture
phase. Of course, lower initial temperature still results in a lower
final temperature, but just in proportion. With nonideal starting
conditions, the upper limit for the cooling factor is determined by
the ratio between the total amount of \textsuperscript{3}He in the
system and the amount of \textsuperscript{3}He in the mixture phase
$c_{\mathrm{F,max}}=\left(n_{3}+n_{\mathrm{m,3}}\right)/n_{\mathrm{m,3}}$.
In conclusion, in order to achieve optimal cooling by the adiabatic
melting method, it is essential to minimize the amount of \textsuperscript{3}He\textendash \textsuperscript{4}He
mixture in the initial state.

Figure~\ref{fig:Cooling-factors} shows not only the cooling factors
in complete melting processes where the final state contains no pure
\textsuperscript{3}He phase, but also the cooling factors for four
different incomplete meltings. The effect on the cooling factor caused
by partial melting is not nearly as substantial as the presence of
the initial mixture phase. From practical aspects, it is not always
desirable to melt the crystal entirely to ensure easier regrowth process
as no new nucleation is required. Also, the experimental cell may
contain surplus of \textsuperscript{3}He to accommodate separate
sintered heat exchanger for the precooling stage, where the mixture
with rather large viscosity is not desired to enter.

\section{Cooling power\label{subsec:Cooling-power}}

The cooling power $\dot{Q}$ of the adiabatic melting process is due
to the latent heat of mixing of \textsuperscript{3}He from the pure
phase to the mixture phase. It is given by
\begin{equation}
\dot{Q}=T\dot{n}_{\mathrm{3}}\left(S_{\mathrm{m,3}}-S_{3}\right),\label{eq:Qdot0}
\end{equation}
where $\dot{n}_{3}$ is the rate at which \textsuperscript{3}He is
transferred between the phases, and $S_{3}$ and $S_{\mathrm{m,}3}$
are the entropies per \textsuperscript{3}He atom in the pure and
the mixture phase, respectively. Well below the $T_{\mathrm{c}}$,
when the temperature is low enough for the \textsuperscript{3}He\textendash \textsuperscript{4}He
mixture to dominate the total entropy of the system, we can ignore
$S_{3}$, and the expression simplifies to
\begin{equation}
\dot{Q}\approx109\,\frac{\mathrm{J}}{\mathrm{mol\,K^{2}}}\,\dot{n}_{3}T^{2}.\label{eq:Qdot1}
\end{equation}
 At $T=100\,\mathrm{\mu K}$, and with $\dot{n}_{3}=100\,\mathrm{\mu mol/s}$,
this gives about $100\,\mathrm{pW}$ of cooling power. To achieve
similar cooling power with an external cooling method, the surface
area of the helium cell would have to be of order $10000\,\mathrm{m^{2}}$,
which was estimated using the thermal boundary resistance values from
Ref.~[\onlinecite{Voncken_sinter}]. Such large surface areas are
hard to obtain in practice, as it would require several kilograms
of sintered silver powder layered on the cell surfaces.

The cooling power depends on the rate $\dot{n}_{3}$, at which \textsuperscript{3}He
atoms transfer from the pure phase to the mixture phase. In the actual
experiment, we cannot directly measure this, but rather we have control
over the extraction rate of \textsuperscript{4}He out from the cell
($\dot{n}_{4}$). In order to facilitate the melting process, the
\textsuperscript{4}He amount corresponding to the molar volume difference
between the solid and liquid phases has to be removed from the cell,
and vice versa if the crystal is grown. Hence it is essential to calculate
the conversion factor $\vartheta$ in $\dot{n}_{3}=\vartheta\dot{n_{4}}$,
which tells us how the extraction rate of \textsuperscript{4}He corresponds
to the phase transfer rate of \textsuperscript{3}He. Let us denote
the total volume of the experimental cell as $v$, and assume that
it contains $n_{3}$ moles of pure \textsuperscript{3}He, $n_{\mathrm{s}}$
moles of solid \textsuperscript{4}He, and $n_{\mathrm{m}}$ moles
of \textsuperscript{3}He\textendash \textsuperscript{4}He mixture,
and thus 
\begin{equation}
v=n_{3}V_{3}+n_{\mathrm{s}}V_{4,\mathrm{s}}+n_{\mathrm{m}}V_{\mathrm{m}},\label{eq:conv1}
\end{equation}
where $V_{3}=26.76\,\mathrm{cm^{3}/mol}$ \citep{Kollar2000a} is
the molar volume of pure \textsuperscript{3}He, and $V_{4,\mathrm{s}}=20.97\,\mathrm{cm^{3}/mol}$
\citep{Driessen1986} is the molar volume of solid \textsuperscript{4}He,
while $V_{\mathrm{m}}$ is as given by Eq.~\eqref{eq:mix-molar-vol}.
When an infinitesimal amount of solid is melted (or grown) the contents
change to
\begin{equation}
\begin{array}{cc}
v= & \left(n_{3}-\mathrm{d}n_{3}\right)V_{3}+\left(n_{\mathrm{s}}-\mathrm{d}n_{\mathrm{s}}\right)V_{4,\mathrm{s}}\\
 & +\left(n_{\mathrm{m}}+\mathrm{d}n_{\mathrm{m}}\right)V_{\mathrm{m}}.
\end{array}\label{eq:conv2}
\end{equation}
Combining Eqs.~\eqref{eq:conv1} and \eqref{eq:conv2}, and taking
the time derivative, results in
\begin{equation}
\dot{n}_{3}V_{3}+\dot{n}_{\mathrm{s}}V_{4,\mathrm{s}}-\dot{n}_{\mathrm{m}}V_{\mathrm{m}}=0.\label{eq:conv2b}
\end{equation}
 The total amount of \textsuperscript{3}He in the system remains
constant, which means that
\begin{equation}
\dot{n}_{3}-x\dot{n}_{\mathrm{m}}=0,\label{eq:conv3}
\end{equation}
while the amount of \textsuperscript{4}He is changing by the amount
required to melt the crystal, giving
\begin{equation}
\dot{n}_{\mathrm{s}}-\left(1-x\right)\dot{n}_{\mathrm{m}}=\dot{n}{}_{4},\label{eq:conv4}
\end{equation}
where $\dot{n}_{4}$ is the rate at which pure \textsuperscript{4}He
is removed from the cell. Using Eqs.~\eqref{eq:conv3} and \eqref{eq:conv4}
together with Eq.~\eqref{eq:conv2b} yields

\begin{equation}
\begin{array}{c}
\vspace{0.1cm}{\displaystyle \dot{n}_{3}V_{3}+\left[\dot{n}_{4}+\left(1-x\right)\frac{\dot{n}_{3}}{x}\right]V_{4,\mathrm{s}}-\dot{n}_{3}V_{\mathrm{m}}=0}\\
\begin{array}{cl}
\Rightarrow\dot{n}_{3} & {\displaystyle =\frac{xV_{4,\mathrm{s}}}{\left(1+\alpha x\right)V_{\mathrm{4,l}}-xV_{3}-\left(1-x\right)V_{4,\mathrm{s}}}\dot{n}_{4}}\\
 & \equiv\vartheta\dot{n}_{4}{\textstyle .}
\end{array}
\end{array}\label{eq:conversion}
\end{equation}

With the numerical values, that can be found from Table~\ref{tab:Parameters},
we get $\vartheta\approx\left(0.84\pm0.01\right)$. Using this, the
low temperature cooling power of the adiabatic melting, expressed
in terms of the \textsuperscript{4}He extraction rate, becomes
\begin{equation}
\dot{Q}=109\,\frac{\mathrm{J}}{\mathrm{mol\,K^{2}}}\,\vartheta\dot{n}_{4}T^{2}\approx91\,\frac{\mathrm{J}}{\mathrm{mol\,K^{2}}}\,\dot{n}_{4}T^{2}.\label{eq:Qdot2}
\end{equation}

\begin{figure}
\includegraphics[width=8.6cm]{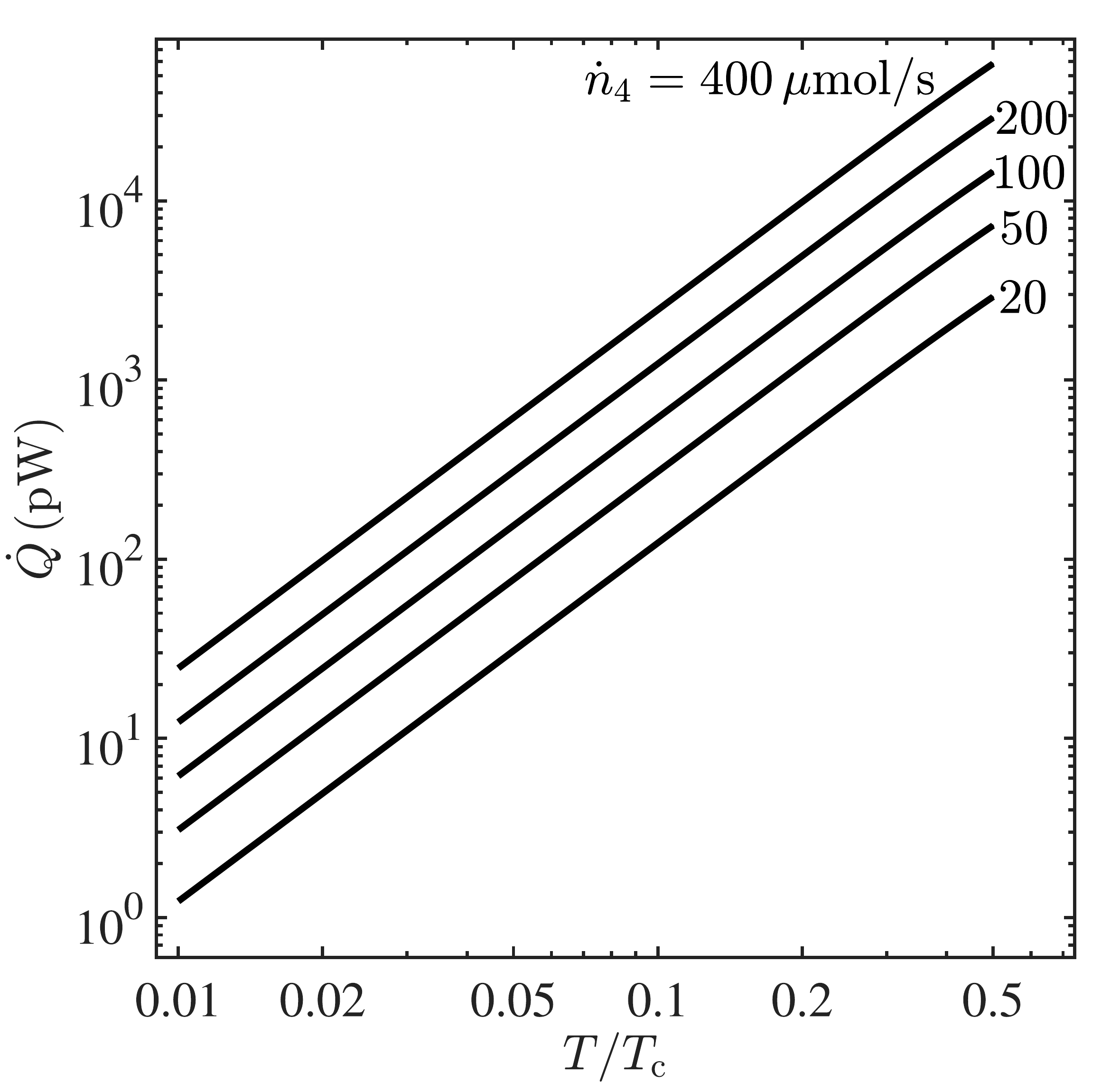}\caption{Cooling power of the adiabatic melting method at different \protect\textsuperscript{4}He
extraction rates $\dot{n}_{4}$ calculated from the entropy difference
between the mixture and pure \protect\textsuperscript{3}He phases
(Eq.~\eqref{eq:Qdot0}).\label{fig:Cooling-power}}
\end{figure}
Figure~\ref{fig:Cooling-power} shows the cooling power up to $400\,\mathrm{\mu mol/s}$
extraction rates. Since there inevitably exists some background heat
leak in any real experimental setup, the lowest possible temperature
is reached when the cooling power matches the heat leak. Furthermore,
the melting process itself may cause rate dependent dissipation due
to the movement of the pure \textsuperscript{3}He\textendash mixture
interface, for example. Therefore there obviously exists an optimal
melting rate which maximizes the cooling power while keeping any additional
losses at sustainable level. With proper cell design the dissipative
losses should not become an issue. $400\,\mathrm{\mu mol/s}$ under
$100\,\mathrm{pW}$ load results in the equilibrium at $T\approx0.022T_{\mathrm{c}}\approx60\,\mathrm{\mu K}$.

As a side note, another useful conversion factor is the change in
the amount of solid in the cell $\dot{n}_{\mathrm{s}}$, when $\dot{n}_{4}$
\textsuperscript{4}He is added or removed. We can solve it by eliminating
$\dot{n}_{3}$, and $\dot{n}_{\mathrm{m}}$ from Eqs.~\eqref{eq:conv3},
\eqref{eq:conv4}, and \eqref{eq:conversion}, yielding

\begin{equation}
\dot{n}_{\mathrm{s}}=\left[1+\left(1-\frac{1}{x}\right)\vartheta\right]\dot{n}_{4}\approx10.5\dot{n}_{4}.
\end{equation}
This is useful in determining the amount of solid \textsuperscript{4}He
in the cell. Further, the denominator of Eq.~\eqref{eq:conversion}
can be rearranged to 
\begin{equation}
\begin{array}{ll}
\Delta V & =\left(V_{4,\mathrm{l}}-V_{4,\mathrm{s}}\right)+x\left(\alpha V_{4,\mathrm{l}}-V_{3}+V_{4,\mathrm{s}}\right)\\
 & \approx\left(2.19-1.99x\right)\,\mathrm{cm^{3}/mol},
\end{array}\label{eq:Delta_V}
\end{equation}
which is the change in the molar volume between solid and liquid phase
in the saturated mixture. The first term is the molar volume difference
between pure liquid \textsuperscript{4}He and solid \textsuperscript{4}He,
while the second term is due to the presence of \textsuperscript{3}He.

\section{Crystallization pressure\label{sec:Crystallization-pressure}}

\begin{figure*}[t]
\includegraphics[width=12.5cm]{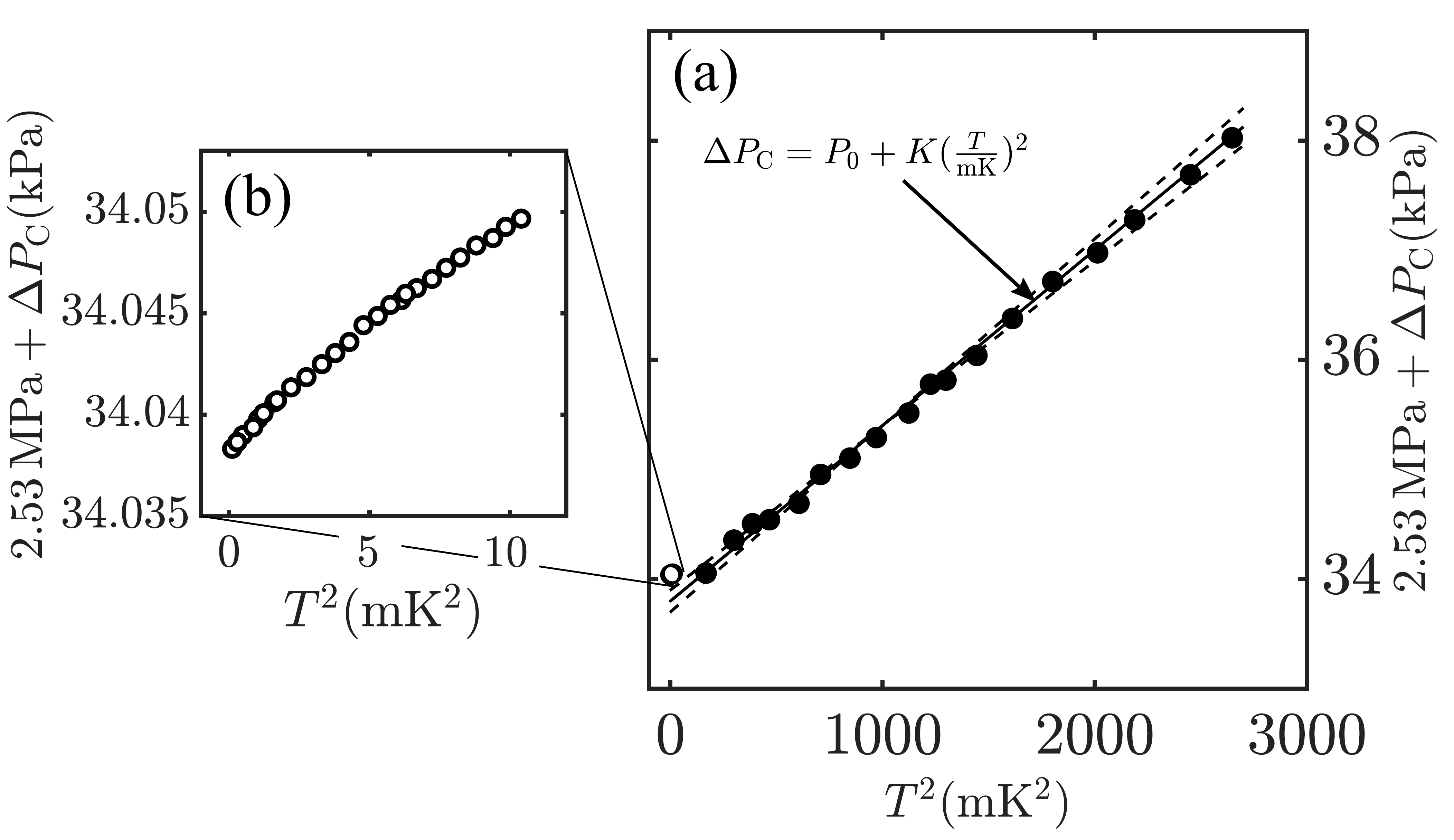}\caption{Difference of the crystallization pressure $\Delta P_{\mathrm{C}}$
from the pure \protect\textsuperscript{4}He zero-temperature value
($2.530\,\mathrm{MPa}$) as a function of $T^{2}$. (a) The datapoints
are the experimental data by Salmela \emph{et al.} \citep{Salmela2011}
($\bullet$) and Pentti \emph{et al.} \citep{Pentti_Thermometry}
($\circ$), while the solid line is a fit to the ($\bullet$) data
with $K=\mathrm{d}P_{\mathrm{C}}/\mathrm{d}\left(T^{2}\right)=\left(1.6\pm0.1\right)\,\mathrm{Pa\,\left(mK\right)^{-2}}$
and $P_{0}=\left(33.8\pm0.1\right)\,\mathrm{kPa}$. The dashed lines
indicate the confidence bounds of the fit. (b) Close-up of the crystallization
pressure data by Pentti \emph{et al.} \citep{Pentti_Thermometry}.\label{fig:Crystallization_pressure}}
\end{figure*}
Once we have the entropies and molar volumes for all components of
the system, we can work out the slope of the crystallization pressure
as the function of temperature through the Clausius-Clapeyron relation
$\mathrm{d}P_{\mathrm{C}}/\mathrm{d}T=\Delta S/\Delta V$. This is
a directly measurable quantity. Both at the zero-temperature limit
and above the $T_{\mathrm{c}}$ up to about $50\,\mathrm{mK}$ this
is directly proportional to temperature, so that the derivative of
$P_{\mathrm{C}}$ with respect to $T^{2}$ is expected to be constant
at these regimes.

Above the $T_{\mathrm{c}}$, the quadratic crystallization pressure
coefficient is given by \citep{Pentti_Thermometry}

\begin{equation}
\begin{array}{cl}
{\displaystyle \frac{\mathrm{d}P_{\mathrm{\mathrm{C}}}}{\mathrm{d}\left(T^{2}\right)}\Bigg|_{T>T_{\mathrm{c}}}} & {\displaystyle =\frac{x\left(S_{\mathrm{m,3}}-S_{3}\right)}{2\Delta V}}\\
 & {\displaystyle =\frac{x\pi^{2}R}{4\Delta V}\left(\frac{1}{T_{\mathrm{F,m}}}-\frac{1}{T_{\mathrm{F},3}}\right),}
\end{array}\label{eq:dPdT2, aboveTc}
\end{equation}
where $\Delta V$ is given by Eq.~\eqref{eq:Delta_V}, and $T_{\mathrm{F,m}}$
and $T_{\mathrm{F,3}}=\pi^{2}(2\gamma)^{-1}$ are the mixture and
the pure \textsuperscript{3}He Fermi temperatures, respectively.
At the zero-temperature limit (effectively when $T\lesssim0.2T_{\mathrm{c}}$),
this reduces to \citep{Pentti_Thermometry}
\begin{equation}
\frac{\mathrm{d}P_{\mathrm{C}}}{\mathrm{d}\left(T^{2}\right)}\Bigg|_{T\rightarrow0}=\frac{xS_{\mathrm{m,3}}}{2\Delta V}=\frac{x\pi^{2}R}{4\Delta VT_{\mathrm{F,m}}},\label{eq:dPdT2, belowTc}
\end{equation}
as $S_{3}\ll S_{\mathrm{m,3}}$. The ratio of the coefficients between
the two regimes is then $\left(1-T_{\mathrm{F,m}}/T_{\mathrm{F,3}}\right)^{-1}$,
which assumes the numerical value 1.36 with our choice of parameters
(Table~\ref{tab:Parameters}).

There are experimental data on these coefficients at different temperature
intervals in Refs.~[\onlinecite{Pentti_Thermometry}], [\onlinecite{Salmela2011}],
and [\onlinecite{Rysti2014}]. Pentti \emph{et al.} \citep{Pentti_Thermometry}
give data within a limited temperature range both below and above
the $T_{\mathrm{c}}$ and quote the values $0.92\,\mathrm{Pa\left(mK\right)^{-2}}$
above the $T_{\mathrm{c}}$ and $1.52\,\mathrm{Pa\,\left(mK\right)^{-2}}$
at the zero-temperature limit, resulting in a ratio 1.65 between the
two. Salmela \emph{et al.} \citep{Salmela2011} give data in the normal
state up to about $50\,\mathrm{mK}$. However, they present fitted
values for the quadratic coefficients only at constant concentrations
below saturation (thus on crucially different two-phase systems),
but the paper also gives data on the saturated system as discrete
points. Performing a similar fit upon these data gives the quadratic
coefficient $\left(1.6\pm0.1\right)\,\mathrm{Pa\,\left(mK\right)^{-2}}$,
represented in Fig.~\ref{fig:Crystallization_pressure}. This result
is in perfect agreement with the corresponding value $\mathrm{d}P_{\mathrm{C}}/\mathrm{d}\left(T^{2}\right)=1.60\,\mathrm{Pa\left(mK\right)^{-2}}$
calculated with Eq.~\eqref{eq:dPdT2, aboveTc} using the parameters
at Table~\ref{tab:Parameters} {[}below the $T_{\mathrm{c}}$ with
Eq.~\eqref{eq:dPdT2, belowTc} we get $2.17\,\mathrm{Pa\,\left(mK\right)^{-2}}${]}.
This result thus supports the validity of our adopted parameter values.
In particular, our choice for the $T_{\mathrm{c}}=2.6\,\mathrm{mK}$
at the mixture crystallization pressure, the overall temperature scale,
and the suggested scaling for the normal state heat capacity coefficient
$\gamma$ of pure \textsuperscript{3}He get some backing from this.
Without scaling the Greywall's\citep{Greywall1986} $\gamma$-value,
the above $T_{\mathrm{c}}$ slope would become $1.48\,\mathrm{Pa\,\left(mK\right)^{-2}}$,
which is also off from the value determined by Pentti \emph{et al.}\citep{Pentti_Thermometry}
{[}$0.92\,\mathrm{Pa\left(mK\right)^{-2}}${]}.

Yet another set of data can be found from Ref.~[\onlinecite{Rysti2014}],
whose measurements extend beyond the domain of validity of the quadratic
temperature dependence. These authors give the quadratic coefficient
as $1.285\,\mathrm{Pa\left(mK\right)^{-2}}$ and find it necessary
to amend the description by a quartic term $-2.065\,\mathrm{MPa\left(mK\right)^{-4}}$,
good from $60\,\mathrm{mK}$ to $140\,\mathrm{mK}$. The deviation
from the quadratic behavior is caused in part by the increase in the
liquid mixture saturation concentration when the temperature rises,
by the molar volumes departing from their constant values, but most
importantly by the fact that \textsuperscript{3}He begins to dissolve
into the solid phase as well at that range of temperatures. The extremely
neat quadratic behavior of $P_{\mathrm{C}}\left(T\right)$ below $50\,\mathrm{mK}$
demonstrates that the solid phase is indeed free from dissolved \textsuperscript{3}He
there.

The inconveniently broad range of the quadratic coefficients above
the $T_{\mathrm{c}}$ indicated by these works is, of course, somewhat
disconcerting. The measurements of Refs.~[\onlinecite{Salmela2011}]
and [\onlinecite{Pentti_Thermometry}] utilized differential pressure
gauges with extremely good sensitivity, but the measurements of Ref.~[\onlinecite{Pentti_Thermometry}]
may have suffered from the effect of rather small reference volume,
as discussed in Ref.~[\onlinecite{Sebedash2006}]. This deficit was
improved by the measurements in Ref.~[\onlinecite{Salmela2011}].
Also, the temperature span covered in Ref.~[\onlinecite{Pentti_Thermometry}]
was rather limited and the quoted value for the saturation concentration
$7.3\%$ is questionable. No good reason for the discrepancy between
the measurements in Refs.~[\onlinecite{Salmela2011}] and [\onlinecite{Rysti2014}]
can be given, except that the necessity to include the quartic term
in the fit of Ref.~[\onlinecite{Rysti2014}] may have introduced
some bias towards the quadratic term as well. The conclusion must
be that the genuine values for these parameters at very low temperatures
are not yet as well established as one might wish.

\section{Conclusions\label{sec:Conclusions}}

Growing the solid phase into a helium mixture at low temperature can
result in a complete phase-separation into solid \textsuperscript{4}He
and liquid \textsuperscript{3}He. Adiabatic melting of such solid
\textsuperscript{4}He, and its following mixing with liquid \textsuperscript{3}He
is a cooling method that can be used in attempts to reach sub-$100\,\mathrm{\mu K}$
temperatures in superfluid \textsuperscript{3}He and saturated \textsuperscript{3}He\textendash \textsuperscript{4}He
mixture at its crystallization pressure. The ability to reach the
lowest possible temperature is strongly dependent on the mixture content
of the experimental volume before the melting process is initiated:
relative to pure superfluid \textsuperscript{3}He, \textsuperscript{3}He\textendash \textsuperscript{4}He
mixture carries a large amount of entropy, and therefore its presence
in the initial state can significantly limit the final temperature.
The ideal initial state would contain only solid \textsuperscript{4}He
and pure \textsuperscript{3}He, and since below the $T_{\mathrm{c}}$,
the pure \textsuperscript{3}He entropy will decrease exponentially,
the total entropy content of the system drops rapidly enabling reduction
in temperature ideally by several orders of magnitude.

The question regarding the practical execution of the experiment is
how to minimize the amount of the initial state mixture. If the mixture
amount is determined by some intrinsic property of the setup, such
as geometry disrupting the growth of the solid, or porous structures
(\emph{e.g.} sinter) trapping the mixture phase, one cannot reduce
it below some threshold value. Another question is whether it is safe
to assume that there are no \textsuperscript{3}He inclusions in the
solid phase. Such \textsuperscript{3}He bubbles in the crystal would
remain hotter than the bulk liquid during the precooling process,
and while melting one should then be able to observe sudden heating
spikes caused by the release of these inclusions. This can obviously
be avoided by proper growing conditions \citep{Pantalei2010}, and
is not expected to be a serious issue.

Another crucial point is to determine where the heat leak to the experimental
cell is coming from, and how to minimize it. Some of it is coming
from the precooling stage through the thermal boundary resistance
bottleneck, since the liquid is cooled to a lower temperature than
the cell structures at the melting period. Measurements themselves
contribute to this and the connecting capillaries are bound to conduct
heat from the hotter parts of the cryostat. Since one can never completely
get rid of the heat leak, an obvious question is, what is the optimal
melting rate of the solid under the given conditions. The heat leak
would have the least effect on the final temperature if the melting
was done as quickly as possible, limited by the critical velocity
in the \textsuperscript{4}He extraction line, or the time needed
for performing the necessary measurements. But if there are some dissipative
losses related to the movement of the pure \textsuperscript{3}He\textendash mixture-interface,
then there may exist possibly lower optimum value. These questions
will be addressed in the future, once our running experiment has produced
sufficient amount of data to enable such analysis.
\begin{acknowledgments}
The authors wish to thank J. Rysti for insightful discussions. This
work was supported by the Jenny and Antti Wihuri Foundation Grant
No. 00170320, and it utilized the facilities provided by Aalto University
at OtaNano - Low Temperature Laboratory infrastructure. 
\end{acknowledgments}

\bibliographystyle{apsrev4-1}

\begin{thebibliography}{35}%
	\makeatletter
	\providecommand \@ifxundefined [1]{%
		\@ifx{#1\undefined}
	}%
	\providecommand \@ifnum [1]{%
		\ifnum #1\expandafter \@firstoftwo
		\else \expandafter \@secondoftwo
		\fi
	}%
	\providecommand \@ifx [1]{%
		\ifx #1\expandafter \@firstoftwo
		\else \expandafter \@secondoftwo
		\fi
	}%
	\providecommand \natexlab [1]{#1}%
	\providecommand \enquote  [1]{``#1''}%
	\providecommand \bibnamefont  [1]{#1}%
	\providecommand \bibfnamefont [1]{#1}%
	\providecommand \citenamefont [1]{#1}%
	\providecommand \href@noop [0]{\@secondoftwo}%
	\providecommand \href [0]{\begingroup \@sanitize@url \@href}%
	\providecommand \@href[1]{\@@startlink{#1}\@@href}%
	\providecommand \@@href[1]{\endgroup#1\@@endlink}%
	\providecommand \@sanitize@url [0]{\catcode `\\12\catcode `\$12\catcode
		`\&12\catcode `\#12\catcode `\^12\catcode `\_12\catcode `\%12\relax}%
	\providecommand \@@startlink[1]{}%
	\providecommand \@@endlink[0]{}%
	\providecommand \url  [0]{\begingroup\@sanitize@url \@url }%
	\providecommand \@url [1]{\endgroup\@href {#1}{\urlprefix }}%
	\providecommand \urlprefix  [0]{URL }%
	\providecommand \Eprint [0]{\href }%
	\providecommand \doibase [0]{http://dx.doi.org/}%
	\providecommand \selectlanguage [0]{\@gobble}%
	\providecommand \bibinfo  [0]{\@secondoftwo}%
	\providecommand \bibfield  [0]{\@secondoftwo}%
	\providecommand \translation [1]{[#1]}%
	\providecommand \BibitemOpen [0]{}%
	\providecommand \bibitemStop [0]{}%
	\providecommand \bibitemNoStop [0]{.\EOS\space}%
	\providecommand \EOS [0]{\spacefactor3000\relax}%
	\providecommand \BibitemShut  [1]{\csname bibitem#1\endcsname}%
	\let\auto@bib@innerbib\@empty
	\bibitem [{\citenamefont {Kapitza}(1938)}]{KAPITZA1938}%
	\BibitemOpen
	\bibfield  {author} {\bibinfo {author} {\bibfnamefont {P.}~\bibnamefont
			{Kapitza}},\ }\href {\doibase 10.1038/141074a0} {\bibfield  {journal}
		{\bibinfo  {journal} {Nature}\ }\textbf {\bibinfo {volume} {141}},\ \bibinfo
		{pages} {74} (\bibinfo {year} {1938})}\BibitemShut {NoStop}%
	\bibitem [{\citenamefont {Allen}\ and\ \citenamefont
		{Misener}(1938)}]{ALLEN1938}%
	\BibitemOpen
	\bibfield  {author} {\bibinfo {author} {\bibfnamefont {J.~F.}\ \bibnamefont
			{Allen}}\ and\ \bibinfo {author} {\bibfnamefont {A.~D.}\ \bibnamefont
			{Misener}},\ }\href {\doibase 10.1038/142643a0} {\bibfield  {journal}
		{\bibinfo  {journal} {Nature}\ }\textbf {\bibinfo {volume} {142}},\ \bibinfo
		{pages} {643} (\bibinfo {year} {1938})}\BibitemShut {NoStop}%
	\bibitem [{\citenamefont {Osheroff}\ \emph
		{et~al.}(1972{\natexlab{a}})\citenamefont {Osheroff}, \citenamefont
		{Richardson},\ and\ \citenamefont {Lee}}]{Osheroff1972}%
	\BibitemOpen
	\bibfield  {author} {\bibinfo {author} {\bibfnamefont {D.~D.}\ \bibnamefont
			{Osheroff}}, \bibinfo {author} {\bibfnamefont {R.~C.}\ \bibnamefont
			{Richardson}}, \ and\ \bibinfo {author} {\bibfnamefont {D.~M.}\ \bibnamefont
			{Lee}},\ }\href {\doibase 10.1103/physrevlett.28.885} {\bibfield  {journal}
		{\bibinfo  {journal} {Phys. Rev. Lett.}\ }\textbf {\bibinfo {volume} {28}},\
		\bibinfo {pages} {885} (\bibinfo {year} {1972}{\natexlab{a}})}\BibitemShut
	{NoStop}%
	\bibitem [{\citenamefont {Osheroff}\ \emph
		{et~al.}(1972{\natexlab{b}})\citenamefont {Osheroff}, \citenamefont {Gully},
		\citenamefont {Richardson},\ and\ \citenamefont {Lee}}]{Osheroff1972a}%
	\BibitemOpen
	\bibfield  {author} {\bibinfo {author} {\bibfnamefont {D.~D.}\ \bibnamefont
			{Osheroff}}, \bibinfo {author} {\bibfnamefont {W.~J.}\ \bibnamefont {Gully}},
		\bibinfo {author} {\bibfnamefont {R.~C.}\ \bibnamefont {Richardson}}, \ and\
		\bibinfo {author} {\bibfnamefont {D.~M.}\ \bibnamefont {Lee}},\ }\href
	{\doibase 10.1103/physrevlett.29.920} {\bibfield  {journal} {\bibinfo
			{journal} {Phys. Rev. Lett.}\ }\textbf {\bibinfo {volume} {29}},\ \bibinfo
		{pages} {920} (\bibinfo {year} {1972}{\natexlab{b}})}\BibitemShut {NoStop}%
	\bibitem [{\citenamefont {Bardeen}\ \emph {et~al.}(1967)\citenamefont
		{Bardeen}, \citenamefont {Baym},\ and\ \citenamefont {Pines}}]{Bardeen1967}%
	\BibitemOpen
	\bibfield  {author} {\bibinfo {author} {\bibfnamefont {J.}~\bibnamefont
			{Bardeen}}, \bibinfo {author} {\bibfnamefont {G.}~\bibnamefont {Baym}}, \
		and\ \bibinfo {author} {\bibfnamefont {D.}~\bibnamefont {Pines}},\ }\href
	{\doibase 10.1103/physrev.156.207} {\bibfield  {journal} {\bibinfo  {journal}
			{Phys. Rev.}\ }\textbf {\bibinfo {volume} {156}},\ \bibinfo {pages} {207}
		(\bibinfo {year} {1967})}\BibitemShut {NoStop}%
	\bibitem [{\citenamefont {Rysti}\ \emph {et~al.}(2012)\citenamefont {Rysti},
		\citenamefont {Tuoriniemi},\ and\ \citenamefont
		{Salmela}}]{Effective_3He_interactions}%
	\BibitemOpen
	\bibfield  {author} {\bibinfo {author} {\bibfnamefont {J.}~\bibnamefont
			{Rysti}}, \bibinfo {author} {\bibfnamefont {J.~T.}\ \bibnamefont
			{Tuoriniemi}}, \ and\ \bibinfo {author} {\bibfnamefont {A.~J.}\ \bibnamefont
			{Salmela}},\ }\href {\doibase 10.1103/PhysRevB.85.134529} {\bibfield
		{journal} {\bibinfo  {journal} {Phys. Rev. B}\ }\textbf {\bibinfo {volume}
			{85}},\ \bibinfo {pages} {134529/1} (\bibinfo {year} {2012})}\BibitemShut
	{NoStop}%
	\bibitem [{\citenamefont {Al-Sugheir}\ \emph {et~al.}(2006)\citenamefont
		{Al-Sugheir}, \citenamefont {Ghassib},\ and\ \citenamefont
		{Joudeh}}]{AL-SUGHEIR2006}%
	\BibitemOpen
	\bibfield  {author} {\bibinfo {author} {\bibfnamefont {M.~K.}\ \bibnamefont
			{Al-Sugheir}}, \bibinfo {author} {\bibfnamefont {H.~B.}\ \bibnamefont
			{Ghassib}}, \ and\ \bibinfo {author} {\bibfnamefont {B.~R.}\ \bibnamefont
			{Joudeh}},\ }\href {\doibase 10.1142/s0217979206034844} {\bibfield  {journal}
		{\bibinfo  {journal} {Int. J. Mod. Phys. B}\ }\textbf {\bibinfo {volume}
			{20}},\ \bibinfo {pages} {2491} (\bibinfo {year} {2006})}\BibitemShut
	{NoStop}%
	\bibitem [{\citenamefont {Sandouqa}\ \emph {et~al.}(2011)\citenamefont
		{Sandouqa}, \citenamefont {Joudeh}, \citenamefont {Al-Sugheir},\ and\
		\citenamefont {Ghassib}}]{Sandouqa2011}%
	\BibitemOpen
	\bibfield  {author} {\bibinfo {author} {\bibfnamefont {A.}~\bibnamefont
			{Sandouqa}}, \bibinfo {author} {\bibfnamefont {B.}~\bibnamefont {Joudeh}},
		\bibinfo {author} {\bibfnamefont {M.}~\bibnamefont {Al-Sugheir}}, \ and\
		\bibinfo {author} {\bibfnamefont {H.}~\bibnamefont {Ghassib}},\ }\href
	{\doibase 10.12693/aphyspola.119.807} {\bibfield  {journal} {\bibinfo
			{journal} {Acta Phys. Pol. A}\ }\textbf {\bibinfo {volume} {119}},\ \bibinfo
		{pages} {807} (\bibinfo {year} {2011})}\BibitemShut {NoStop}%
	\bibitem [{\citenamefont {Ferrier-Barbut}\ \emph {et~al.}(2014)\citenamefont
		{Ferrier-Barbut}, \citenamefont {Delehaye}, \citenamefont {Laurent},
		\citenamefont {Grier}, \citenamefont {Pierce}, \citenamefont {Rem},
		\citenamefont {Chevy},\ and\ \citenamefont
		{Salomon}}]{Bose_Fermi_superfluids}%
	\BibitemOpen
	\bibfield  {author} {\bibinfo {author} {\bibfnamefont {I.}~\bibnamefont
			{Ferrier-Barbut}}, \bibinfo {author} {\bibfnamefont {M.}~\bibnamefont
			{Delehaye}}, \bibinfo {author} {\bibfnamefont {S.}~\bibnamefont {Laurent}},
		\bibinfo {author} {\bibfnamefont {A.~T.}\ \bibnamefont {Grier}}, \bibinfo
		{author} {\bibfnamefont {M.}~\bibnamefont {Pierce}}, \bibinfo {author}
		{\bibfnamefont {B.~S.}\ \bibnamefont {Rem}}, \bibinfo {author} {\bibfnamefont
			{F.}~\bibnamefont {Chevy}}, \ and\ \bibinfo {author} {\bibfnamefont
			{C.}~\bibnamefont {Salomon}},\ }\href@noop {} {\bibfield  {journal} {\bibinfo
			{journal} {Science}\ }\textbf {\bibinfo {volume} {345}},\ \bibinfo {pages}
		{1035} (\bibinfo {year} {2014})}\BibitemShut {NoStop}%
	\bibitem [{\citenamefont {Roy}\ \emph {et~al.}(2017)\citenamefont {Roy},
		\citenamefont {Green}, \citenamefont {Bowler},\ and\ \citenamefont
		{Gupta}}]{Roy2017}%
	\BibitemOpen
	\bibfield  {author} {\bibinfo {author} {\bibfnamefont {R.}~\bibnamefont
			{Roy}}, \bibinfo {author} {\bibfnamefont {A.}~\bibnamefont {Green}}, \bibinfo
		{author} {\bibfnamefont {R.}~\bibnamefont {Bowler}}, \ and\ \bibinfo {author}
		{\bibfnamefont {S.}~\bibnamefont {Gupta}},\ }\href@noop {} {\bibfield
		{journal} {\bibinfo  {journal} {Phys. Rev. Lett.}\ }\textbf {\bibinfo
			{volume} {118}},\ \bibinfo {pages} {055301} (\bibinfo {year}
		{2017})}\BibitemShut {NoStop}%
	\bibitem [{\citenamefont {Oh}\ \emph {et~al.}(1994)\citenamefont {Oh},
		\citenamefont {Ishimoto}, \citenamefont {Kawae}, \citenamefont {Nakagawa},
		\citenamefont {Ishikawa}, \citenamefont {Hata}, \citenamefont {Kodama},\ and\
		\citenamefont {Ikehata}}]{Oh1994}%
	\BibitemOpen
	\bibfield  {author} {\bibinfo {author} {\bibfnamefont {G.~H.}\ \bibnamefont
			{Oh}}, \bibinfo {author} {\bibfnamefont {Y.}~\bibnamefont {Ishimoto}},
		\bibinfo {author} {\bibfnamefont {T.}~\bibnamefont {Kawae}}, \bibinfo
		{author} {\bibfnamefont {M.}~\bibnamefont {Nakagawa}}, \bibinfo {author}
		{\bibfnamefont {O.}~\bibnamefont {Ishikawa}}, \bibinfo {author}
		{\bibfnamefont {T.}~\bibnamefont {Hata}}, \bibinfo {author} {\bibfnamefont
			{T.}~\bibnamefont {Kodama}}, \ and\ \bibinfo {author} {\bibfnamefont
			{S.}~\bibnamefont {Ikehata}},\ }\href {\doibase 10.1007/bf00751787}
	{\bibfield  {journal} {\bibinfo  {journal} {J. Low Temp. Phys.}\ }\textbf
		{\bibinfo {volume} {95}},\ \bibinfo {pages} {525} (\bibinfo {year}
		{1994})}\BibitemShut {NoStop}%
	\bibitem [{\citenamefont {Sebedash}\ \emph {et~al.}(2007)\citenamefont
		{Sebedash}, \citenamefont {Tuoriniemi}, \citenamefont {Boldarev},
		\citenamefont {Pentti},\ and\ \citenamefont {Salmela}}]{Adiabatic_Melting}%
	\BibitemOpen
	\bibfield  {author} {\bibinfo {author} {\bibfnamefont {A.~P.}\ \bibnamefont
			{Sebedash}}, \bibinfo {author} {\bibfnamefont {J.~T.}\ \bibnamefont
			{Tuoriniemi}}, \bibinfo {author} {\bibfnamefont {S.~T.}\ \bibnamefont
			{Boldarev}}, \bibinfo {author} {\bibfnamefont {E.~M.~M.}\ \bibnamefont
			{Pentti}}, \ and\ \bibinfo {author} {\bibfnamefont {A.~J.}\ \bibnamefont
			{Salmela}},\ }\href {\doibase 10.1007/s10909-007-9443-5} {\bibfield
		{journal} {\bibinfo  {journal} {J. Low Temp. Phys.}\ }\textbf {\bibinfo
			{volume} {148}},\ \bibinfo {pages} {725} (\bibinfo {year}
		{2007})}\BibitemShut {NoStop}%
	\bibitem [{\citenamefont {Sebedash}\ \emph {et~al.}(2017)\citenamefont
		{Sebedash}, \citenamefont {Boldarev}, \citenamefont {Riekki},\ and\
		\citenamefont {Tuoriniemi}}]{Sebedash_QFS}%
	\BibitemOpen
	\bibfield  {author} {\bibinfo {author} {\bibfnamefont {A.}~\bibnamefont
			{Sebedash}}, \bibinfo {author} {\bibfnamefont {S.}~\bibnamefont {Boldarev}},
		\bibinfo {author} {\bibfnamefont {T.}~\bibnamefont {Riekki}}, \ and\ \bibinfo
		{author} {\bibfnamefont {J.}~\bibnamefont {Tuoriniemi}},\ }\href {\doibase
		10.1007/s10909-017-1755-5} {\bibfield  {journal} {\bibinfo  {journal} {J. Low
				Temp. Phys}\ }\textbf {\bibinfo {volume} {187}},\ \bibinfo {pages} {588}
		(\bibinfo {year} {2017})}\BibitemShut {NoStop}%
	\bibitem [{\citenamefont {Sebedash}(1997)}]{Sebedash1997}%
	\BibitemOpen
	\bibfield  {author} {\bibinfo {author} {\bibfnamefont {A.~P.}\ \bibnamefont
			{Sebedash}},\ }\href {\doibase 10.1134/1.567360} {\bibfield  {journal}
		{\bibinfo  {journal} {JETP Lett.}\ }\textbf {\bibinfo {volume} {65}},\
		\bibinfo {pages} {276} (\bibinfo {year} {1997})}\BibitemShut {NoStop}%
	\bibitem [{\citenamefont {Pentti}\ \emph {et~al.}(2007)\citenamefont {Pentti},
		\citenamefont {Tuoriniemi}, \citenamefont {Salmela},\ and\ \citenamefont
		{Sebedash}}]{Pentti_Thermometry}%
	\BibitemOpen
	\bibfield  {author} {\bibinfo {author} {\bibfnamefont {E.}~\bibnamefont
			{Pentti}}, \bibinfo {author} {\bibfnamefont {J.}~\bibnamefont {Tuoriniemi}},
		\bibinfo {author} {\bibfnamefont {A.}~\bibnamefont {Salmela}}, \ and\
		\bibinfo {author} {\bibfnamefont {A.}~\bibnamefont {Sebedash}},\ }\href
	{\doibase 10.1007/s10909-006-9267-8} {\bibfield  {journal} {\bibinfo
			{journal} {J. Low Temp. Phys.}\ }\textbf {\bibinfo {volume} {146}},\ \bibinfo
		{pages} {71} (\bibinfo {year} {2007})}\BibitemShut {NoStop}%
	\bibitem [{\citenamefont {Dobbs}(2001)}]{Dobbs2001}%
	\BibitemOpen
	\bibfield  {author} {\bibinfo {author} {\bibfnamefont {E.~R.}\ \bibnamefont
			{Dobbs}},\ }\href@noop {} {\emph {\bibinfo {title} {Helium Three
				(International Series of Monographs on Physics)}}}\ (\bibinfo  {publisher}
	{Oxford University Press},\ \bibinfo {year} {2001})\BibitemShut {NoStop}%
	\bibitem [{\citenamefont {Pantalei}\ \emph {et~al.}(2010)\citenamefont
		{Pantalei}, \citenamefont {Rojas}, \citenamefont {Edwards}, \citenamefont
		{Maris},\ and\ \citenamefont {Balibar}}]{Pantalei2010}%
	\BibitemOpen
	\bibfield  {author} {\bibinfo {author} {\bibfnamefont {C.}~\bibnamefont
			{Pantalei}}, \bibinfo {author} {\bibfnamefont {X.}~\bibnamefont {Rojas}},
		\bibinfo {author} {\bibfnamefont {D.~O.}\ \bibnamefont {Edwards}}, \bibinfo
		{author} {\bibfnamefont {H.~J.}\ \bibnamefont {Maris}}, \ and\ \bibinfo
		{author} {\bibfnamefont {S.}~\bibnamefont {Balibar}},\ }\href {\doibase
		10.1007/s10909-010-0159-6} {\bibfield  {journal} {\bibinfo  {journal} {J. Low
				Temp. Phys.}\ }\textbf {\bibinfo {volume} {159}},\ \bibinfo {pages} {452}
		(\bibinfo {year} {2010})}\BibitemShut {NoStop}%
	\bibitem [{\citenamefont {Balibar}\ \emph {et~al.}(2000)\citenamefont
		{Balibar}, \citenamefont {Mizusaki},\ and\ \citenamefont
		{Sasaki}}]{Balibar2000}%
	\BibitemOpen
	\bibfield  {author} {\bibinfo {author} {\bibfnamefont {S.}~\bibnamefont
			{Balibar}}, \bibinfo {author} {\bibfnamefont {T.}~\bibnamefont {Mizusaki}}, \
		and\ \bibinfo {author} {\bibfnamefont {Y.}~\bibnamefont {Sasaki}},\ }\href
	{\doibase 10.1023/a:1004669102741} {\bibfield  {journal} {\bibinfo  {journal}
			{J. Low Temp. Phys.}\ }\textbf {\bibinfo {volume} {120}},\ \bibinfo {pages}
		{293} (\bibinfo {year} {2000})}\BibitemShut {NoStop}%
	\bibitem [{\citenamefont {Balibar}(2002)}]{Balibar2002}%
	\BibitemOpen
	\bibfield  {author} {\bibinfo {author} {\bibfnamefont {S.}~\bibnamefont
			{Balibar}},\ }\href {\doibase 10.1023/a:1021412529571} {\bibfield  {journal}
		{\bibinfo  {journal} {J. Low Temp. Phys.}\ }\textbf {\bibinfo {volume}
			{129}},\ \bibinfo {pages} {363} (\bibinfo {year} {2002})}\BibitemShut
	{NoStop}%
	\bibitem [{\citenamefont {Pentti}\ \emph {et~al.}(2008)\citenamefont {Pentti},
		\citenamefont {Tuoriniemi}, \citenamefont {Salmela},\ and\ \citenamefont
		{Sebedash}}]{Pentti_etal_solubility}%
	\BibitemOpen
	\bibfield  {author} {\bibinfo {author} {\bibfnamefont {E.~M.}\ \bibnamefont
			{Pentti}}, \bibinfo {author} {\bibfnamefont {J.~T.}\ \bibnamefont
			{Tuoriniemi}}, \bibinfo {author} {\bibfnamefont {A.~J.}\ \bibnamefont
			{Salmela}}, \ and\ \bibinfo {author} {\bibfnamefont {A.~P.}\ \bibnamefont
			{Sebedash}},\ }\href {\doibase 10.1103/PhysRevB.78.064509} {\bibfield
		{journal} {\bibinfo  {journal} {Phys. Rev. B}\ }\textbf {\bibinfo {volume}
			{78}},\ \bibinfo {pages} {064509} (\bibinfo {year} {2008})}\BibitemShut
	{NoStop}%
	\bibitem [{\citenamefont {Rusby}\ \emph {et~al.}(2002)\citenamefont {Rusby},
		\citenamefont {Durieux}, \citenamefont {Reesink}, \citenamefont {Hudson},
		\citenamefont {Schuster}, \citenamefont {K\"uhne}, \citenamefont {Fogle},
		\citenamefont {Soulen},\ and\ \citenamefont {Adams}}]{Rusby2002}%
	\BibitemOpen
	\bibfield  {author} {\bibinfo {author} {\bibfnamefont {R.~L.}\ \bibnamefont
			{Rusby}}, \bibinfo {author} {\bibfnamefont {M.}~\bibnamefont {Durieux}},
		\bibinfo {author} {\bibfnamefont {A.~L.}\ \bibnamefont {Reesink}}, \bibinfo
		{author} {\bibfnamefont {R.~P.}\ \bibnamefont {Hudson}}, \bibinfo {author}
		{\bibfnamefont {G.}~\bibnamefont {Schuster}}, \bibinfo {author}
		{\bibfnamefont {M.}~\bibnamefont {K\"uhne}}, \bibinfo {author} {\bibfnamefont
			{W.~E.}\ \bibnamefont {Fogle}}, \bibinfo {author} {\bibfnamefont {R.~J.}\
			\bibnamefont {Soulen}}, \ and\ \bibinfo {author} {\bibfnamefont {E.~D.}\
			\bibnamefont {Adams}},\ }\href {\doibase 10.1023/a:1013791823354} {\bibfield
		{journal} {\bibinfo  {journal} {J. Low Temp. Phys.}\ }\textbf {\bibinfo
			{volume} {126}},\ \bibinfo {pages} {633} (\bibinfo {year}
		{2002})}\BibitemShut {NoStop}%
	\bibitem [{\citenamefont {Manninen}\ \emph {et~al.}(2016)\citenamefont
		{Manninen}, \citenamefont {Ranni}, \citenamefont {Rysti}, \citenamefont
		{Todoshchenko},\ and\ \citenamefont {Tuoriniemi}}]{Manninen2016}%
	\BibitemOpen
	\bibfield  {author} {\bibinfo {author} {\bibfnamefont {M.~S.}\ \bibnamefont
			{Manninen}}, \bibinfo {author} {\bibfnamefont {A.}~\bibnamefont {Ranni}},
		\bibinfo {author} {\bibfnamefont {J.}~\bibnamefont {Rysti}}, \bibinfo
		{author} {\bibfnamefont {I.~A.}\ \bibnamefont {Todoshchenko}}, \ and\
		\bibinfo {author} {\bibfnamefont {J.~T.}\ \bibnamefont {Tuoriniemi}},\ }\href
	{\doibase 10.1007/s10909-016-1590-0} {\bibfield  {journal} {\bibinfo
			{journal} {J. Low Temp. Phys.}\ }\textbf {\bibinfo {volume} {183}},\ \bibinfo
		{pages} {399} (\bibinfo {year} {2016})}\BibitemShut {NoStop}%
	\bibitem [{\citenamefont {Lifshitz}\ and\ \citenamefont
		{Pitaevskii}(1980)}]{Lifshitz1980}%
	\BibitemOpen
	\bibfield  {author} {\bibinfo {author} {\bibfnamefont {E.}~\bibnamefont
			{Lifshitz}}\ and\ \bibinfo {author} {\bibfnamefont {L.~P.}\ \bibnamefont
			{Pitaevskii}},\ }\href
	{https://www.amazon.com/Statistical-Physics-Theory-Condensed-State/dp/0750626364?SubscriptionId=AKIAIOBINVZYXZQZ2U3A&tag=chimbori05-20&linkCode=xm2&camp=2025&creative=165953&creativeASIN=0750626364}
	{\emph {\bibinfo {title} {Statistical Physics: Theory of the Condensed State
				(Pt 2)}}}\ (\bibinfo  {publisher} {Butterworth-Heinemann},\ \bibinfo {year}
	{1980})\BibitemShut {NoStop}%
	\bibitem [{\citenamefont {Greywall}(1986)}]{Greywall1986}%
	\BibitemOpen
	\bibfield  {author} {\bibinfo {author} {\bibfnamefont {D.~S.}\ \bibnamefont
			{Greywall}},\ }\href {\doibase 10.1103/physrevb.33.7520} {\bibfield
		{journal} {\bibinfo  {journal} {Phys. Rev. B}\ }\textbf {\bibinfo {volume}
			{33}},\ \bibinfo {pages} {7520} (\bibinfo {year} {1986})}\BibitemShut
	{NoStop}%
	\bibitem [{\citenamefont {Alvesalo}\ \emph {et~al.}(1981)\citenamefont
		{Alvesalo}, \citenamefont {Haavasoja},\ and\ \citenamefont
		{Manninen}}]{Alvesalo1981}%
	\BibitemOpen
	\bibfield  {author} {\bibinfo {author} {\bibfnamefont {T.~A.}\ \bibnamefont
			{Alvesalo}}, \bibinfo {author} {\bibfnamefont {T.}~\bibnamefont {Haavasoja}},
		\ and\ \bibinfo {author} {\bibfnamefont {M.~T.}\ \bibnamefont {Manninen}},\
	}\href {\doibase 10.1007/bf00655140} {\bibfield  {journal} {\bibinfo
			{journal} {J. Low Temp. Phys.}\ }\textbf {\bibinfo {volume} {45}},\ \bibinfo
		{pages} {373} (\bibinfo {year} {1981})}\BibitemShut {NoStop}%
	\bibitem [{\citenamefont {Todoschenko}\ \emph {et~al.}(2002)\citenamefont
		{Todoschenko}, \citenamefont {Alles}, \citenamefont {Babkin}, \citenamefont
		{Parshin},\ and\ \citenamefont {Tsepelin}}]{Todoschenko2002}%
	\BibitemOpen
	\bibfield  {author} {\bibinfo {author} {\bibfnamefont {I.~A.}\ \bibnamefont
			{Todoschenko}}, \bibinfo {author} {\bibfnamefont {H.}~\bibnamefont {Alles}},
		\bibinfo {author} {\bibfnamefont {A.}~\bibnamefont {Babkin}}, \bibinfo
		{author} {\bibfnamefont {A.~Y.}\ \bibnamefont {Parshin}}, \ and\ \bibinfo
		{author} {\bibfnamefont {V.}~\bibnamefont {Tsepelin}},\ }\href {\doibase
		10.1023/a:1014259906576} {\bibfield  {journal} {\bibinfo  {journal} {J. Low
				Temp. Phys.}\ }\textbf {\bibinfo {volume} {126}},\ \bibinfo {pages} {1449}
		(\bibinfo {year} {2002})}\BibitemShut {NoStop}%
	\bibitem [{\citenamefont {Serene}\ and\ \citenamefont
		{Rainer}(1983)}]{Serene1983}%
	\BibitemOpen
	\bibfield  {author} {\bibinfo {author} {\bibfnamefont {J.}~\bibnamefont
			{Serene}}\ and\ \bibinfo {author} {\bibfnamefont {D.}~\bibnamefont
			{Rainer}},\ }\href {\doibase 10.1016/0370-1573(83)90051-0} {\bibfield
		{journal} {\bibinfo  {journal} {Phys. Rep.}\ }\textbf {\bibinfo {volume}
			{101}},\ \bibinfo {pages} {221} (\bibinfo {year} {1983})}\BibitemShut
	{NoStop}%
	\bibitem [{\citenamefont {Tanaka}\ \emph {et~al.}(2000)\citenamefont {Tanaka},
		\citenamefont {Hatakeyama}, \citenamefont {Noma},\ and\ \citenamefont
		{Satoh}}]{Tanaka2000}%
	\BibitemOpen
	\bibfield  {author} {\bibinfo {author} {\bibfnamefont {E.}~\bibnamefont
			{Tanaka}}, \bibinfo {author} {\bibfnamefont {K.}~\bibnamefont {Hatakeyama}},
		\bibinfo {author} {\bibfnamefont {S.}~\bibnamefont {Noma}}, \ and\ \bibinfo
		{author} {\bibfnamefont {T.}~\bibnamefont {Satoh}},\ }\href {\doibase
		10.1016/s0011-2275(00)00052-7} {\bibfield  {journal} {\bibinfo  {journal}
			{Cryogenics}\ }\textbf {\bibinfo {volume} {40}},\ \bibinfo {pages} {365}
		(\bibinfo {year} {2000})}\BibitemShut {NoStop}%
	\bibitem [{\citenamefont {Watson}\ \emph {et~al.}(1969)\citenamefont {Watson},
		\citenamefont {Reppy},\ and\ \citenamefont
		{Richardson}}]{Watson_Reppy_Richardson}%
	\BibitemOpen
	\bibfield  {author} {\bibinfo {author} {\bibfnamefont {G.~E.}\ \bibnamefont
			{Watson}}, \bibinfo {author} {\bibfnamefont {J.~D.}\ \bibnamefont {Reppy}}, \
		and\ \bibinfo {author} {\bibfnamefont {R.~C.}\ \bibnamefont {Richardson}},\
	}\href {\doibase 10.1103/PhysRev.188.384} {\bibfield  {journal} {\bibinfo
			{journal} {Phys. Rev.}\ }\textbf {\bibinfo {volume} {188}},\ \bibinfo {pages}
		{384} (\bibinfo {year} {1969})}\BibitemShut {NoStop}%
	\bibitem [{\citenamefont {Voncken}\ \emph {et~al.}(1996)\citenamefont
		{Voncken}, \citenamefont {Riese}, \citenamefont {Roobol}, \citenamefont
		{Konig},\ and\ \citenamefont {Pobell}}]{Voncken_sinter}%
	\BibitemOpen
	\bibfield  {author} {\bibinfo {author} {\bibfnamefont {A.~P.~J.}\
			\bibnamefont {Voncken}}, \bibinfo {author} {\bibfnamefont {D.}~\bibnamefont
			{Riese}}, \bibinfo {author} {\bibfnamefont {L.~P.}\ \bibnamefont {Roobol}},
		\bibinfo {author} {\bibfnamefont {R.}~\bibnamefont {Konig}}, \ and\ \bibinfo
		{author} {\bibfnamefont {F.}~\bibnamefont {Pobell}},\ }\href {\doibase
		10.1007/BF00754629} {\bibfield  {journal} {\bibinfo  {journal} {J. Low Temp.
				Phys.}\ }\textbf {\bibinfo {volume} {105}},\ \bibinfo {pages} {93} (\bibinfo
		{year} {1996})}\BibitemShut {NoStop}%
	\bibitem [{\citenamefont {Kollar}\ and\ \citenamefont
		{Vollhardt}(2000)}]{Kollar2000a}%
	\BibitemOpen
	\bibfield  {author} {\bibinfo {author} {\bibfnamefont {M.}~\bibnamefont
			{Kollar}}\ and\ \bibinfo {author} {\bibfnamefont {D.}~\bibnamefont
			{Vollhardt}},\ }\href {\doibase 10.1103/physrevb.61.15347} {\bibfield
		{journal} {\bibinfo  {journal} {Phys. Rev. B}\ }\textbf {\bibinfo {volume}
			{61}},\ \bibinfo {pages} {15347} (\bibinfo {year} {2000})}\BibitemShut
	{NoStop}%
	\bibitem [{\citenamefont {Driessen}\ \emph {et~al.}(1986)\citenamefont
		{Driessen}, \citenamefont {van~der Poll},\ and\ \citenamefont
		{Silvera}}]{Driessen1986}%
	\BibitemOpen
	\bibfield  {author} {\bibinfo {author} {\bibfnamefont {A.}~\bibnamefont
			{Driessen}}, \bibinfo {author} {\bibfnamefont {E.}~\bibnamefont {van~der
				Poll}}, \ and\ \bibinfo {author} {\bibfnamefont {I.~F.}\ \bibnamefont
			{Silvera}},\ }\href {\doibase 10.1103/physrevb.33.3269} {\bibfield  {journal}
		{\bibinfo  {journal} {Phys. Rev. B}\ }\textbf {\bibinfo {volume} {33}},\
		\bibinfo {pages} {3269} (\bibinfo {year} {1986})}\BibitemShut {NoStop}%
	\bibitem [{\citenamefont {Salmela}\ \emph {et~al.}(2011)\citenamefont
		{Salmela}, \citenamefont {Sebedash}, \citenamefont {Rysti}, \citenamefont
		{Pentti},\ and\ \citenamefont {Tuoriniemi}}]{Salmela2011}%
	\BibitemOpen
	\bibfield  {author} {\bibinfo {author} {\bibfnamefont {A.}~\bibnamefont
			{Salmela}}, \bibinfo {author} {\bibfnamefont {A.}~\bibnamefont {Sebedash}},
		\bibinfo {author} {\bibfnamefont {J.}~\bibnamefont {Rysti}}, \bibinfo
		{author} {\bibfnamefont {E.}~\bibnamefont {Pentti}}, \ and\ \bibinfo {author}
		{\bibfnamefont {J.}~\bibnamefont {Tuoriniemi}},\ }\href@noop {} {\bibfield
		{journal} {\bibinfo  {journal} {Phys. Rev. B}\ }\textbf {\bibinfo {volume}
			{83}},\ \bibinfo {pages} {134510} (\bibinfo {year} {2011})}\BibitemShut
	{NoStop}%
	\bibitem [{\citenamefont {Rysti}\ \emph {et~al.}(2014)\citenamefont {Rysti},
		\citenamefont {Manninen},\ and\ \citenamefont {Tuoriniemi}}]{Rysti2014}%
	\BibitemOpen
	\bibfield  {author} {\bibinfo {author} {\bibfnamefont {J.}~\bibnamefont
			{Rysti}}, \bibinfo {author} {\bibfnamefont {M.~S.}\ \bibnamefont {Manninen}},
		\ and\ \bibinfo {author} {\bibfnamefont {J.}~\bibnamefont {Tuoriniemi}},\
	}\href {\doibase 10.1007/s10909-014-1154-0} {\bibfield  {journal} {\bibinfo
			{journal} {J. Low Temp. Phys.}\ }\textbf {\bibinfo {volume} {175}},\ \bibinfo
		{pages} {739} (\bibinfo {year} {2014})}\BibitemShut {NoStop}%
	\bibitem [{\citenamefont {Sebedash}\ \emph {et~al.}(2006)\citenamefont
		{Sebedash}, \citenamefont {Tuoriniemi}, \citenamefont {Boldarev},
		\citenamefont {Pentti},\ and\ \citenamefont {Salmela}}]{Sebedash2006}%
	\BibitemOpen
	\bibfield  {author} {\bibinfo {author} {\bibfnamefont {A.}~\bibnamefont
			{Sebedash}}, \bibinfo {author} {\bibfnamefont {J.~T.}\ \bibnamefont
			{Tuoriniemi}}, \bibinfo {author} {\bibfnamefont {S.}~\bibnamefont
			{Boldarev}}, \bibinfo {author} {\bibfnamefont {E.~M.}\ \bibnamefont
			{Pentti}}, \ and\ \bibinfo {author} {\bibfnamefont {A.~J.}\ \bibnamefont
			{Salmela}},\ }\href@noop {} {\bibfield  {journal} {\bibinfo  {journal} {{AIP}
				Conf. Proc.}\ } (\bibinfo {year} {2006})}\BibitemShut {NoStop}%
\end{thebibliography}

\end{document}